\documentclass[onecolumn,12pt]{article}
\usepackage{jheppub}
\usepackage{ifpdf}
\usepackage{graphicx}
\usepackage{amsfonts}
\usepackage{amsmath}
\usepackage{amssymb}
\usepackage{epsfig}
\usepackage{pdfpages}
\usepackage{graphicx,epstopdf}
\usepackage{caption}
\usepackage[position=b]{subcaption}
\usepackage[makeroom]{cancel}
\usepackage{hyperref}
\hypersetup{colorlinks=true,allcolors=blue} 
\usepackage{array}

\def\al{\alpha}

\def\ga{\gamma}

\def\th{\vartheta}

\def\om{\omega}

\def\Om{\Omega}

\def\m{\mathcal}

\def\cal{\calligra}

\def\hor{\big|_{\mathcal{N}}}

\def\csform{\Theta}

\newcommand{\comment}[1]{{\bf *******{#1}*******}}

\newcommand{\wh}[1]{\widehat{#1}}
\def\IZ {\mathbb{Z}}
\newcommand{\beq}{\begin{equation}}
\newcommand{\eeq}{\end{equation}}
\newcommand{\bea}{\begin{eqnarray}}
\newcommand{\eea}{\end{eqnarray}}
\newcommand{\nn}{\nonumber}

\begin{document}
	\baselineskip=15.5pt
	\baselineskip 20 pt
	\pagestyle{plain}
	\setcounter{page}{1}
	
	
	\def\del{{\partial}}
	\def\vev#1{\left\langle #1 \right\rangle}
	\def\cn{{\cal N}}
	\def\co{{\cal O}}
	\def\IC{{\mathbb C}}
	\def\IR{{\mathbb R}}
	\def\IZ{{\mathbb Z}}
	\def\RP{{\bf RP}}
	\def\CP{{\bf CP}}
	\def\Poincare{{Poincar\'e }}
	\def\tr{{\rm tr\,}}
	\def\tp{{\tilde \Phi}}

	\def\TL{\hfil$\displaystyle{##}$}
	\def\TR{$\displaystyle{{}##}$\hfil}
	\def\TC{\hfil$\displaystyle{##}$\hfil}
	\def\TT{\hbox{##}}
	\def\HLINE{\noalign{\vskip1\jot}\hline\noalign{\vskip1\jot}}
	\def\seqalign#1#2{\vcenter{\openup1\jot
			\halign{\strut #1\cr #2 \cr}}}
	\def\lbldef#1#2{\expandafter\gdef\csname #1\endcsname {#2}}
	\def\eqn#1#2{\lbldef{#1}{(\ref{#1})}
		\begin{equation} #2 \label{#1} \end{equation}}
	\def\eqalign#1{\vcenter{\openup1\jot
			\halign{\strut\span\TL & \span\TR\cr #1 \cr
	}}}
	\def\eno#1{(\ref{#1})}
	\def\href#1#2{#2}
	\def\half{{\frac 1 2}}

	\def\ads{{\it AdS}}
	\def\adsp{{\it AdS}$_{p+2}$}
	\def\cft{{\it CFT}}
	
	\newcommand{\ber}{\begin{eqnarray}}
	\newcommand{\eer}{\end{eqnarray}}
	
	\newcommand{\prt}[1]{{\left( {#1} \right)}}
	
	\newcommand{\beqar}{\begin{eqnarray}}
	\newcommand{\cN}{{\cal N}}
	\newcommand{\cO}{{\cal O}}
	\newcommand{\cA}{{\cal A}}
	\newcommand{\cT}{{\cal T}}
	\newcommand{\cF}{{\cal F}}
	\newcommand{\cC}{{\cal C}}
	\newcommand{\cR}{{\cal R}}
	\newcommand{\cW}{{\cal W}}
	\newcommand{\eeqar}{\end{eqnarray}}
	\newcommand{\tht}{\thteta}
	\newcommand{\lm}{\lambda}\newcommand{\Lm}{\Lambda}
	\newcommand{\eps}{\epsilon}
	

	\newcommand{\nonu}{\nonumber}
	\newcommand{\oh}{\displaystyle{\frac{1}{2}}}
	\newcommand{\dsl}
	{\kern.06em\hbox{\raise.15ex\hbox{$/$}\kern-.56em\hbox{$\partial$}}}
	\newcommand{\id}{i\!\!\not\!\partial}
	\newcommand{\as}{\not\!\! A}
	\newcommand{\ps}{\not\! p}
	\newcommand{\ks}{\not\! k}
	\newcommand{\D}{{\cal{D}}}
	\newcommand{\dv}{d^2x}
	\newcommand{\Z}{{\cal Z}}
	\newcommand{\N}{{\cal N}}
	\newcommand{\Dsl}{\not\!\! D}
	\newcommand{\Bsl}{\not\!\! B}
	\newcommand{\Psl}{\not\!\! P}
	\newcommand{\eeqarr}{\end{eqnarray}}
\newcommand{\ZZ}{{\rm \kern 0.275em Z \kern -0.92em Z}\;}

\def\azb{A_{\bar z}} \def\az{A_z} \def\bzb{B_{\bar z}} \def\bz{B_z}
\def\czb{C_{\bar z}} \def\cz{C_z} \def\dzb{D_{\bar z}} \def\dz{D_z}
\def\im{{\hbox{\rm Im}}} \def\mod{{\hbox{\rm mod}}} \def\tr{{\hbox{\rm Tr}}}
\def\ch{{\hbox{\rm ch}}} \def\imp{{\hbox{\sevenrm Im}}}
\def\trp{{\hbox{\sevenrm Tr}}} \def\vol{{\hbox{\rm Vol}}}
\def\rl{\Lambda_{\hbox{\sevenrm R}}} \def\wl{\Lambda_{\hbox{\sevenrm W}}}
\def\fc{{\cal F}_{k+\cox}} \def\vev{vacuum expectation value}
\def\nodiv{\mid{\hbox{\hskip-7.8pt/}}}
\def\ie{{\em i.e.}}
\def\ie{\hbox{\it i.e.}}
\def\HWA{\vskip3mm\centerline{\comment{Here we are}}\vskip3mm}

\def\CC{{\mathchoice
	{\rm C\mkern-8mu\vrule height1.45ex depth-.05ex
		width.05em\mkern9mu\kern-.05em}
	{\rm C\mkern-8mu\vrule height1.45ex depth-.05ex
		width.05em\mkern9mu\kern-.05em}
	{\rm C\mkern-8mu\vrule height1ex depth-.07ex
		width.035em\mkern9mu\kern-.035em}
	{\rm C\mkern-8mu\vrule height.65ex depth-.1ex
		width.025em\mkern8mu\kern-.025em}}}

\def\RR{{\rm I\kern-1.6pt {\rm R}}}
\def\NN{{\rm I\!N}}
\def\ZZ{{\rm Z}\kern-3.8pt {\rm Z} \kern2pt}
\def\IB{\relax{\rm I\kern-.18em B}}
\def\ID{\relax{\rm I\kern-.18em D}}
\def\II{\relax{\rm I\kern-.18em I}}
\def\IP{\relax{\rm I\kern-.18em P}}
\newcommand{\CS}{{\scriptstyle {\rm CS}}}
\newcommand{\CSs }{{\scriptscriptstyle {\rm CS}}}
\newcommand{\rc}{\nonumber\\}
\newcommand{\bear}{\begin{eqnarray}}
\newcommand{\eear}{\end{eqnarray}}
\newcommand{\W}{{\cal W}}
\newcommand{\F}{{\cal F}}
\newcommand{\x}{{\cal O}}
\newcommand{\LL}{{\cal L}}

\def\mani{{\cal M}}
\def\cO{{\cal O}}
\def\calb{{\cal B}}
\def\calw{{\cal W}}
\def\calz{{\cal Z}}
\def\cald{{\cal D}}
\def\calc{{\cal C}}
\def\to{\rightarrow}
\def\ele{{\hbox{\sevenrm L}}}
\def\ere{{\hbox{\sevenrm R}}}
\def\zb{{\bar z}}
\def\wb{{\bar w}}
\def\nodiv{\mid{\hbox{\hskip-7.8pt/}}}
\def\menos{\hbox{\hskip-2.9pt}}
\def\dr{\dot R_}
\def\drr{\dot r_}
\def\ds{\dot s_}
\def\da{\dot A_}
\def\dga{\dot \gamma_}
\def\ga{\gamma_}
\def\dal{\dot\alpha_}
\def\al{\alpha_}
\def\cl{{closed}}
\def\cls{{closing}}
\def\vev{vacuum expectation value}
\def\tr{{\rm tr\,}}
\def\to{\rightarrow}
\def\too{\longrightarrow}


\def\a{\alpha}
\def\b{\beta}
\def\c{\gamma}
\def\d{\delta}
\def\e{\epsilon}           
\def\f{\phi}               
\def\vf{\varphi}  \def\tvf{\tilde{\varphi}}
\def\vp{\varphi}
\def\g{\gamma}
\def\h{\eta}
\def\j{\psi}
\def\k{\kappa}                    
\def\l{\lambda}
\def\m{\mu}
\def\n{\nu}
\def\o{\omega}  \def\w{\omega}
\def\q{\th}  \def\th{\theta}                  
\def\r{\rho}                                     
\def\s{\sigma}                                   
\def\t{\tau}
\def\u{\upsilon}
\def\x{\xi}
\def\z{\zeta}
\def\pt{\tilde{\varphi}}
\def\tt{\tilde{\theta}}
\def\lab{\label}
\def\6{\partial}
\def\wg{\wedge}
\def\atanh{{\rm arctanh}}
\def\bpsi{\bar{\psi}}
\def\bt{\bar{\theta}}
\def\bvf{\bar{\varphi}}

\def\dd{\mathrm{d}}
\def\na{\nabla}
\def\eq{&=&}
\def\f2{\phi_2}
\def\ff{\phi_2\phi_2}
\newcommand{\pref}[1]{\prt{\ref{#1}}}
\newcommand{\bh}[1]{\widehat{\bar{#1}}}

\def\dh{\widehat{\phi}}
\def\gh{\widehat{g}}
\def\gfh{\widehat{g}_{44}}

\def\dg{\sqrt{-g}}
\def\dr{\mathrm{det}}
\def\dgtr{\mathrm{det}\, g^{tr} }
\def\hgtr{\mathrm{det}\, \widehat{g}^{tr} }
\def\gff{g_{44}} 
\def\gt{\tilde{g}}
\def\tg{\tilde{g}}
\def\gzz{g_{zz}}
\def\gyy{G_{\psi\psi}}
\def\tgzz{\tilde{g_{zz}}}
\def\ozz{\omega^2_{zz}}
\def\oyy{\omega^2_{yy}}

\def\om{\Omega} \def \Om{\Omega}
\def\tom{\tilde{\omega}}
\def\tOm{\tilde{\Omega}}
\def\of{\omega^2_{44}}
\def\ob{\Omega^2_{44}}

\def\wf {\widehat{\phi_2}}
\def\wff{\widehat{\phi_2} \widehat{\phi_2}}
\def\W2{\omega^2}
\def\Wt{\tilde{\omega}}
\def\wttt{\tilde{\omega}^{2\;t}_t}
\def\wtrr{\tilde{\omega}^{2\;r}_r}

\def\tb{\tilde{b}}
\def\hb{\widehat{b}}
\def\Ht{\tilde{H}}
\def \rp{r_+}

\def\dgt{\delta \tilde{g}}
\def\lzz{L^0_{\;0}}
\def\loo{L^1_{\;1}}
\def\ltt{L^2_{\;2}}

\def\hlzz{\wh{L}^0_{\;0}}
\def\hloo{\wh{L}^1_{\;1}}
\def\hltt{\wh{L}^2_{\;2}}
\def\bb{{||}}
\def\thi{{\th^{(i)}}}
\def\thj{{\th^{(j)}}}
\def\G{\Gamma}
\def\hor{|_\mathcal{N}}
\def\inb{|_\mathcal{B}}
\def\bif{\mathcal B}
\def \mn {\mathcal{N}}

\def\gh{\widehat{g}}
\def\gfh{\widehat{g}_{44}}
\def\dh{\widehat{\phi}}
\def \Gh{\widehat G}

\def\gm{\gamma_{-}}
\def\es{e^\s}
\def \ems{e^{-\s}}
\def \eds{e^{2\s}}
\def \emds{e^{-2\s}}
\def \upsi {\underline{\psi}}
\def \bm{{\bar \m}}
\def \bn{{\bar \n}}
\def \mr{\mathring}

\def\gp{\gamma_{+}}

\def\cH{\mathcal{H}}
\def\dLLz{\delta_{\Gamma}}
\def\dLLx{\delta_{\xi}}
\def\lambdaE{(\lambda^E_\xi)}
\def\lambdaEp{(\lambda^{E'}_\xi)}
\def\lambdaEz{(\lambda^E_\zeta)}
\def\lambdaEpz{(\lambda^{E'}_\zeta)}
\def\LieD{\mathcal{L}_{\xi}}
\def\LieDz{\mathcal{L}_{\zeta}} 

\def\m{\mu}
\def\n{\nu}
\def\o{\omega}  \def\w{\omega}
\def\csform{\Theta}

\def \txi{\tilde{\xi}}

\def\bus{\mathbb B}
\def\li     {\mathcal{L}_\xi}
\def\litxi {\mathcal{L}_{\tilde{\xi}}}
\def\ko{\mathcal{K}_\xi}
\def \DD{\mathrm{D}}
\def \ii{\mathrm{i}_\xi}
\def \alv{\bar \delta_\xi^e }
\def \lxi{\lambda_\xi^e} 
\def\mh{\mathcal{H}}
\def\ixi{\mathrm i_\xi}
\def\lxie {\lambda_\xi^E} 
\def \alve{\bar \delta_\xi^E }
\def\lxieab {(\lambda_\xi^E)_A{}^B} 
\def\lxieba {(\lambda_\xi^E)_B{}^A} 
\def \kxie{\mathcal K_\xi^E}
\def \kxieab{(\mathcal K_\xi^E)_A{}^B}
\def \kxieba{(\mathcal K_\xi^E)_B{}^A}

\def\bh {\bar H}
\def\wm { \Omega^{(-)}}
\def\wp  { \Omega^{(+)} }
\def \wmpsi{ \Omega^{(-)}_\psi }
\def \wppsi{ \Omega^{(+)}_\psi }
\def \hwmpsi{ \widehat \Omega^{(-)}  _\psi }
\def \hwppsi{  \widehat  \Omega^{(+)}_\psi }
\def\cm {C_\mu}
\def\cn {C_\nu}
\def\gpp {G_{\psi\psi}}
\def\ogpp{\frac{1}{G_{\psi \psi }}}

\def \ta { \breve }
\def \cal {\mathcal }
\def \ml {\mathcal L} 
\def \wt {\widetilde}
\def \mc {\mathcal }

%

\newfont{\namefont}{cmr10}
\newfont{\addfont}{cmti7 scaled 1440}
\newfont{\boldmathfont}{cmbx10}
\newfont{\headfontb}{cmbx10 scaled 1728}
\setcounter{equation}{0}

\title{T-duality equivalences beyond string theory}

\author[a]{Jos\'e D. Edelstein,}
\author[b]{Konstantinos Sfetsos,}
\author[c]{J. Anibal Sierra-Garcia,}
\author[a]{Alejandro Vilar L\'opez}

\affiliation[a]{Departamento de F\'isica de Part\'iculas $\&$ Instituto Galego de F\'isica de Altas Enerx\'ias (IGFAE), Universidad de Santiago de Compostela, E-15782 Santiago de Compostela, Spain}
\affiliation[b]{Department of Nuclear and Particle Physics, Faculty of Physics, National and Kapodistrian University of Athens, Athens 15784, Greece}
\affiliation[c]{Department of Physics, Faculty of Science, Chulalongkorn University, Bangkok 10330, Thailand}
\abstract{We examine a two parameter family of gravitational actions which contains higher-derivative terms. These are such that the entire action is invariant under corrected T-duality rules, which we derive explicitly. Generically this action does not describe low energy string backgrounds except for isolated choices for the parameters. Nevertheless, we demonstrate that in this theory the entropy and the temperature of generic non-extremal black hole solutions are T-duality invariant. This further supports the idea put forward in our previous work that T-duality might provide physical equivalences beyond the realm of string theory.}
\date{\today}

\dedicated{Dedicated to the memory of Stephen Hawking, one year after his passing}

\maketitle

\emailAdd{jose.edelstein@usc.es}
\emailAdd{ksfetsos@phys.uoa.gr}
\emailAdd{anibal.sierra.garcia@gmail.com}
\emailAdd{alejandro.vilar.lopez@gmail.com}

\numberwithin{equation}{section}

\section{Introduction}

T-duality was born as an equivalence between string theories in different target spaces.
Geometrically distinct spacetimes (with different background fields) turn out to be physically equivalent solutions of a given string theory. An intriguing aspect of this duality comes from the fact that dual solutions can have fairly different geometric properties; for instance, it is not guaranteed that T-duality maps black hole geometries into other black hole spacetimes. Since properties such as the entropy or temperature of spacetimes possessing horizons \cite{Bekenstein,Hawking} are related to geometric features of the solution, it is a priori unknown whether T-duality is going to respect them or not. This puzzle was answered in the affirmative by analyzing black hole solutions in the NS-NS sector of string theory \cite{Horowitz:1993wt}. Even though the geometry is significantly affected by a T-duality transformation, horizons are mapped into horizons, the entropy and temperature remaining invariant.

Things become less clear when higher-derivative corrections are introduced since, for instance, the entropy ceases to be given by the event horizon area. Do black hole horizons, their entropy and surface gravity, remain invariant under T-duality when these corrections are included? One may be tempted to answer that this is guaranteed by the very fact that these corrections to the low-energy effective action arise from a sigma model, and T-duality is an exact discrete symmetry associated to its target space. In fact, contrary to generic higher-derivative quantum field theories ---no matter how rich their particle content may be---, string theory is thought to possess more symmetries such as the one that tells us that the physics at very small scales cannot be distinguished from that at large scales.

In spite of this observation, albeit T-duality constrains the possible higher-derivative terms in the action \cite{Hohm2015}, there is still room for at least a two-parameter family of four-derivative T-dual invariant theories \cite{MarquesNunez} ---building up on earlier work \cite{Hohm2014}--- which includes but goes beyond String Theory. This brings about a possible additional puzzle: what is the effect of T-duality when acting on the non-stringy black hole members of this family? Does T-duality invariance of, say, their entropy and temperature hold only for those black holes solving the equations of motion of low-energy string theory? It is natural to expect the sigma model origin of the latter to be a crucial aspect behind the result. In particular, given the expectation that the entropy accounts for all the constituent microscopic degrees of freedom, both perturbative and non-perturbative. For that same reason one might expect the counting to fail in a theory belonging to the swampland, much in the same way as those theories are afflicted by other issues such as causality violation \cite{CEMZ}.

We studied this problem in an earlier paper \cite{ESSV1}, in the restricted context of three-dimensional gravity and BTZ black holes. We showed that both the entropy and the temperature of the black holes are unexpectedly invariant under T-duality for the whole two-parameter family, to leading order in the derivative expansion weighted by the inverse mass scale $M_\star^{-2}$. The AdS/CFT correspondence, though, enforces quantization conditions on the parameters. The exceptional feature of three-dimensional gravity together with the exactness of the BTZ solution simplified significantly the analysis. In this paper we aim at completing the task and showing that those results are completely general; {\it i.e.}, valid for black holes in the higher dimensional case too.

Let us be a bit more explicit on the theoretical context where our result is derived. Within the framework of so-called Double Field Theory \cite{Hull:2009mi}, a very fruitful formalism allowing to build low-energy effective actions which are symmetric under T-duality by construction, Marqu\'es and N\'u\~nez \cite{MarquesNunez} recently wrote a two-parameter family of theories governed by the generalized Bergshoeff-de Roo \cite{BergshoeffdeRoo} action:
\begin{eqnarray}
\mathcal{I}_{\rm BdR} &=& \int \dd^{D-1}x \sqrt{-{G}}\, e^{- 2 \Phi} \bigg[ R - 2Ê\Lambda + 4 ( \nabla_M \nabla^M \Phi- \nabla_M \Phi \nabla^M \Phi ) - \frac{1}{12} H'_{MNR}\, H'^{MNR} \nonumber \\ [0.5em]
& & \qquad + \frac{1}{8} \sum_{k=\pm} a_k\, R^{(k)}_{M N A}{}^B R^{(k) M N}{}_B{}^A \bigg] ~ , \label{BergshoeffdeRoo}
\end{eqnarray}
where the sum runs over two signs, $k = \pm$, and the parameters are going to be dubbed $a_- \equiv a$ and $a_+ \equiv b$. Notice that we added a cosmological term to the action presented in \cite{MarquesNunez} ---for free, it is T-dual invariant on its own---, and we work in units where $16 \pi G = 1$. We have further defined:
\begin{equation}
H'_{M N R} = H_{M N R} - \frac{3}{2} \left( a \, \Theta^{(-)}_{M N R} - b \, \Theta^{(+)}_{M N R} \right) ~,
\label{barh}
\end{equation}
$\Theta^{(\pm)}_{M N R}$ being the gravitational Chern-Simons forms
\begin{equation}
\Theta^{(\pm)}_{M N R} = \Omega^{(\pm)}_{[MA}{}^B \partial_N \Omega^{(\pm)}_{R]B}{}^A + \frac{2}{3} \Omega^{(\pm)}_{[MA}{}^B \Omega^{(\pm)}_{NB}{}^C \Omega^{(\pm)}_{R]C}{}^A ~,
\label{ThetaMNR}
\end{equation}
of the pair of torsionful connections:
\begin{equation}
\Omega^{(\pm)}_{MA}{}^B := \Omega_{MA}{}^B \pm \frac{H_{MA}{}^B}{2} ~, 
\label{TorsionnedConnectionDefinition}
\end{equation}
where $\Omega_{MA}{}^B$ is the spin connection and $H_{MA}{}^B = E^N{}_A E^{RB} H_{MNR}$, indices being raised (and lowered) with the vielbein $E^M{}_A$. It is convenient to introduce the 1-form,
\begin{equation}
\cH_{A}{}^B := H_{MA}{}^B\, dx^M ~,
\label{H1form}
\end{equation}
for later purposes. The Riemann tensors, $R^{(\pm)}_{ M N A}{}^{B}$, are also built from the torsionful connections,
\begin{equation}
R^{(\pm)}_{ M N A}{}^{B} = \partial_M \Omega^{(\pm)}_{NA}{}^B - \partial_N \Omega^{(\pm)}_{MA}{}^B + \Omega^{(\pm)}_{MA}{}^C \Omega^{(\pm)}_{NC}{}^B - \Omega^{(\pm)}_{NA}{}^C \Omega^{(\pm)}_{MC}{}^B ~.
\label{RMNAB}
\end{equation}
It is important to realize that we work in a perturbative framework assuming our parameters $a$ and $b$ to be order $M_{*}^{-2}$, and therefore the quadratic (in $a$ and $b$) terms appearing in the previous action are just a convenient form of writing it and they must not be taken into consideration. The part of the action \eqref{BergshoeffdeRoo} which does not contain the perturbative parameters corresponds to the action governing the universal massless NS-NS sector, where $\Phi$ is the dilaton and $B_{MN}$ the Kalb-Ramond two-form potential, which appears through its curvature $H_{MNR}$. For specific values of $a$ and $b$, the first order corrections can be seen to arise in the low-energy effective actions of string theories:
\begin{eqnarray}
& a = b = - \alpha' ~, \qquad \qquad \qquad & {\rm bosonic} ~, \nonumber \\ [0.4em]
& ~a = - \alpha' ~ , \quad b = 0 ~, \qquad \qquad & {\rm heterotic} ~, \label{stringvalues} \\ [0.4em]
& a = b = 0 ~, \qquad \qquad \qquad \quad & {\rm type~II} ~. \nonumber
\end{eqnarray}
The case $a + b = 0$ is also special \cite{HSZ}. However, for generic values of $a$ and $b$ not included in the previous cases we do not know of any sigma model or CFT which could give rise to the generalized Bergshoeff-de Roo action (\ref{BergshoeffdeRoo}). In spite of this, the theory is invariant under T-duality corrected rules whose precise form will be presented later on, provided that we neglect quadratic terms in $a$ and $b$. For the sake of completeness, let us present the equations of motion derived from \eqref{BergshoeffdeRoo}, already obtained in \cite{ESSV1}:
\begin{eqnarray}
& & R - 2 \Lambda + 4 ( \nabla^2 \Phi - (\nabla\Phi)^2 ) - \frac{1}{12} H'_{MNR}\, H'^{MNR} + \frac{1}{8} \sum_{k=\pm} a_k R^{(k)}_{M N A}{}^B R^{(k) MN}{}_B{}^A = 0 ~, \nonumber \\ [0.6em]
& & \nabla_{M} \left[  e^{-2\Phi}  H'^{MNR} + \frac{3}{2} \sum_{k=\pm} a_k \left( e^{-2\Phi} H^{ST[M}  R^{(k) RN]}_{ST} - k \nabla^{(k)}_{S} \left[ e^{-2\Phi} {R^{(k) S[MNR]}} \right] \right) \right] = 0 ~, \nonumber \\ [0.6em]
& & R_{MN} + 2 \nabla_M \nabla_N \Phi - \frac{1}{4} H'_{MRS} H'_{N}{}^{RS} - \frac{1}{4} \sum_{k=\pm} a_k \left[ R^{(k)}_{MRST} R^{(k) RST}_N \right. \label{BREoM} \\ [0.5em]
& & \qquad\quad \left. + e^{2\Phi} \left( 2 {G}_{S(M|} \nabla_R + k H_{RS(M|} \right) \left( {\delta_U}^{S} \nabla^{(k)}_T + k {H_{TU}}^{S} \right) \left( e^{-2\Phi} {R^{(k) TUR}}_{|N)} \right) \right] = 0 ~, \nonumber
\end{eqnarray}
where $\nabla^2 = \nabla_M \nabla^M$, $ (\nabla\Phi)^2 =  \nabla_M \Phi \nabla^M \Phi$, and $\nabla^{(k)}$ is the covariant derivative involving the connection with torsion $\Gamma_{M N}^{(\pm)}{}^R=\Gamma_{M N}^R \mp \frac{1}{2} H_{M N}{}^R$. The (anti)symmetrization is always normalized with the factorial of the number of indices, for instance: $v_{(A} w_{B)} := \frac{1}{2!} (v_A w_B + v_B w_A)$.

The action \eqref{BergshoeffdeRoo} contains explicitly the gravitational Chern-Simons forms $\Theta^{(\pm)}$, and as a consequence it is not Lorentz invariant in general. It can be shown to be invariant under an \emph{anomalous} local Lorentz transformation \cite{MarquesNunez}:
\begin{equation}
\begin{array}{l}
\delta_\Lambda E_M{}^A = E_M{}^B \Lambda_B{}^A ~, \\ [1.0em]
\delta_\Lambda B_{MN} = - \displaystyle\frac{a}{2}\,\partial_{[M} \Lambda_A{}^B \Omega^{(-)}_{N] B}{}^A + \displaystyle\frac{b}{2}\,\partial_{[M} \Lambda_A{}^B \Omega^{(+)}_{N] B}{}^A ~,
\end{array}
\label{AnomalousLorentzTransformation}
\end{equation}
where $\Lambda_A{}^B$ is the infinitesimal parameter. For the heterotic string case, $b=0$, this symmetry transformation is the consequence of the anomaly cancellation \`a la Green-Schwarz, while for the bosonic case $a = b$ we can avoid the necessity of this anomalous transformation through a field redefinition \cite{MarquesNunez}. Note that, despite the word ``anomalous'', the symmetry is exact to linear order in $a$ and $b$; we will use the notation $\mathcal{O}(a, b)$. It will prove to be of primary importance when we discuss entropy of black hole solutions in this theory, and it forces us to consider the vielbein $E_M{}^A$, the two-form $B_{MN}$, and the dilaton $\Phi$ as the basic degrees of freedom.

The paper is structured as follows. Section \ref{SectionCorrectedRules} contains a brief review of the leading order Buscher rules implementing T-duality when $a = b = 0$, after which we present the corrected rules, which constitute a symmetry transformation of the previous action. Section \ref{SectionEntropy} contains the discussion concerning the derivation of the entropy for any solution of the theory containing a bifurcate Killing horizon, which has to take into account all the symmetries present. The entropy turns out to be anomalous Lorentz invariant, as expected. Finally, Section \ref{SectionInvariance} proves in a fairly general situation the invariance of the entropy and temperature associated with a horizon under T-duality. Particularly convenient coordinates and vielbein must be introduced in a neighborhood of the horizon, and this is discussed before showing the actual invariance of the thermodynamic quantities associated with it.

In order to avoid distracting the reader with technical side details, several appendices complement the main text providing all those which are necessary and relevant to follow the more involved calculations. In particular, Appendix \ref{AppendixOtherVielbeins} discusses how the corrected T-duality rules are obtained from the Double Field Theory formalism, and Appendix \ref{AppendixEntropyComputation} provides the complete derivation of the entropy formula. Appendices \ref{AppendixStationarityOfTheDual} and \ref{AppendixAreaInvariance} prove some key technical results needed to show the entropy invariance under T-duality; namely, the fact that the dual of a stationary solution is itself stationary (in the sense required by the entropy derivation) and the invariance of the horizon area under corrected T-duality rules. Finally, in Appendix \ref{AppendixFieldRedefinition} we present an independent check of our results by means of a series of field redefinitions bringing the action to a frame first presented in \cite{MetsaevTseytlin}.

\section{Corrected T-duality rules}
\label{SectionCorrectedRules}

In this section we explain how to apply T-duality in the generalized Bergshoeff-de Roo action \eqref{BergshoeffdeRoo}. We start by reviewing the standard Buscher rules for the set of fields $G_{M N}, B_{M N}$ and $\Phi$. Then we introduce a convenient class of vielbeins, $E_M{}^A$, in order to apply T-duality transformations and discuss the rules in the absence of higher-derivative corrections ---{\it i.e.}, when $a = b = 0$---, using dimensional reduction. We work out the results for the Lorentz connection $\Omega_{M A}{}^B$, which finally allow us to compute the $(a,b)$-corrected T-dual background given by the fields $\wh E_M{}^A, \wh B_{M N}$ and $\wh \Phi$. 

Let us start with the review of leading order results to specify our conventions. The uncorrected rules for $E_M{}^A, B_{M N}$ and $\Phi$ are\footnote{In order to obtain the T-dual of the vielbein, a doubling procedure must be invoked on general grounds; {\it i.e.}, a pair of vielbeins must be introduced (see appendix \ref{AppendixOtherVielbeins} for details). In particular, such procedure is relevant in the derivation of the $(a,b)$-corrected rules. We have included only one of the two dual vielbeins at this point since it is sufficient in the case of the generalized Bergshoeff-de Roo action we are dealing with. This set of rules can be derived from Double Field Theory (see \cite{MarquesNunez,Aldazabal:2013sca}), and also directly in $D+1$ dimensions (see \cite{Hassan:1999bv} and references therein).}
\begin{eqnarray}
\wt E_\m{}^A  &=& E_\m{}^A - \frac{Q_{\psi \m}}{G_{\psi\psi}} E_\psi{}^A ~, \qquad \wt E_\psi{}^A = \frac{E_\psi{}^A}{G_{\psi\psi}} ~, \label{BuscherRulesE}\nn \\ [0.4em]
\wt B_{\psi \m}  \eq - \frac{G_{\psi \m}}{\gyy} ~, \qquad \wt B_{\m\n} = B_{\m\n} - \frac{G_{\psi \m}B_{\psi \n} - B_{\psi \m}G_{\psi \n}}{\gyy} ~, \label{BuscherRulesB} \\ [0.6em]
e^{-2\wt \Phi} \eq e^{-2\Phi} \gyy ~,
\nn 
\label{BuscherRulesPhi}
\end{eqnarray}
where we denoted the fields obtained with these standard uncorrected Buscher rules by a tilde, and $Q_{M N} := G_{M N} + B_{M N}$. Then the transformation of the metric is given by
\begin{eqnarray}
\wt G_{\psi\psi} \eq \frac{1}{G_{\psi\psi}} ~, \qquad \wt G_{\psi \m} = - \frac{B_{\psi \m}}{\gyy} ~, \label{BuscherRulesGpsi}
\\ [0.7em]
\wt G_{\m\n}  \eq G_{\m\n}- \frac{G_{\psi \m} G_{\psi \n} -B_{\psi \m} B_{\psi \n}}{\gyy} ~.
\label{BuscherRulesGmn}
\end{eqnarray}
Our spacetime is $(D+1)$-dimensional, with coordinates $\{x^\m,\psi\}$, where $\psi$ is the coordinate adapted to the U(1) symmetry we T-dualize with respect to, and $\m$ runs over the other $D$ coordinates. The T-dual fields in the presence of non-vanishing values of $a$ and/or $b$ will be denoted by $\widehat\Psi$, where $\Psi$ stands for a configuration of the fundamental fields, namely $\Psi = \{ E_M{}^A, B_{M N}, \Phi\}$. We note that $\widehat\Psi \neq \wt \Psi$, even if $\wt \Psi$ may generically have corrections linear in $a$ and $b$.\footnote{For instance, when tilde acts on $(a,b)$-corrected field configurations, we must keep those corrections in the result; {\it e.g.}, if $G_{\psi\psi} = \frac 1 r + \frac a {r^3}$ then $\wt G_{\psi \psi} = r - \frac a r$ instead of $\wt G_{\psi \psi} = r$. Still, $\widehat G_{\psi\psi} \neq \wt G_{\psi \psi}$, as we will shortly see.} 

Let us now review the leading order transformation of the vielbein and Lorentz connection using the results of \cite{KaloperMeissner} in the particularly convenient ansatz consistent with the U(1) symmetry
\begin{eqnarray}
\dd s^2 \eq (\dd x^\m e_\m{}^a)^2 + e^{2\s} \prt{\dd x^\m V_\m  + \dd\psi}^2 ~, \nn \\ [0.55em]
B \eq b + \half W \wg V + W \wg \dd\psi ~, \label{KMAnsatz} \\ [0.3em]
\Phi \eq \phi + \half \s ~, \nn
\end{eqnarray}
where $a$ runs from 0 to $D-1$. This expression defines (up to $D$-dimensional Lorentz transformations of the vielbein $e_\m{}^a$) the reduced fields $e_\m{}^a, V_\m, b_{\m \n}, W_\m, \phi$ and $\s$. Notice the peculiar reduction of $B$, in which one would normally omit the term $\half W \wg V$. The $(D+1)$-dimensional vielbein is chosen to be $E^A = \{E^a, E^{\underline \psi}\}$, with
\begin{equation} 
E^a = \dd x^\m\, e_\m{}^a ~, \qquad E^{\upsi} = e^{\s} (\dd x^\m\, V_\m + \dd\psi) ~.
\label{KMVielbein}
\end{equation}
It is convenient to define at this point a reduced field strength $h_{\m \n \r}$ \cite{KaloperMeissner}, which \emph{is not the field strength} of $ b_{\m\n}$,
\begin{equation} 
h_{\m\n\r} := 3 \nabla_{[\m} b_{\n\r]} - \frac{3}{2} W_{[\m\n} V_{\r]} - \frac{3}{2} V_{[\m\n} W_{\r]} ~,
\label{ReducedFieldStrength}
\end{equation}
where $V_{\m \n} = \partial_\m V_\n - \partial_\n V_\m$ and $W_{\m \n} = \partial_\m W_\n  - \partial_\n W_\m$ are the usual field strengths of $V_\m$ and $W_\m$. In all $(D+1)$-dimensional quantities, the flat components are referred to the vielbein $ E_M{}^A$. Similarly, in the $D$-dimensional quantities, they are referred to the $D$-dimensional vielbein, $e_\m{}^a$. The decomposition \eqref{KMAnsatz} allows to write the uncorrected Buscher rules \eqref{BuscherRulesE} and \eqref{BuscherRulesGmn} simply as:
\begin{equation}
\wt V_\m = W_\m ~, \qquad \wt W_\m = V_\m ~, \qquad \wt \s = -\s ~,
\label{ReducedRules}
\end{equation}
while $e_\m{}^a, b_{\m \n}$ and $\phi$ remain the same \cite{KaloperMeissner}. In particular, only the component $E^{\underline \psi}$ of the vielbein is modified.

We need to obtain the leading order transformation of the (torsionful) Lorentz connection under T-duality in order to later find their $(a,b)$-corrected rules. These can be achieved by writing $\Om^{(\pm)}_{M A}{}^B$ in terms of reduced fields that then are transformed as in \eqref{ReducedRules}. In other words, we perform the dimensional reduction of $\Om_{A B C} = E_A{}^M \Om_{M B C}$,
\begin{equation}
\Om_{a b c} = \o_{a b c} ~, \qquad \Om_{a b \upsi} = \frac{e^\s V_{a b}} 2 = - \Om_{\upsi a b} ~, \qquad \Om_{\upsi a \upsi} = - \partial_a \s ~,
\label{lorentzconnectionrule}
\end{equation}
where $\o_{a b c}$ is the Lorentz connection of $e_\m{}^a$, $V_{ab} = e_a{}^\m e_b{}^\n V_{\m \n}$, and $\partial_a \s = e_a{}^\m \partial_\m \s$. From these expressions we obtain
\begin{equation}
\begin{array}{l}
\Om^{(\pm)}_{a b c} = \o_{a b c} \pm \displaystyle\half h_{abc} ~, \qquad \Om^{(\pm)}_{\upsi a \upsi} = - \partial_a \s ~, \\ [1.2em]
\Om^{(\pm)} _{a b  \upsi} = \displaystyle\frac{1}{2} \prt{\es V_{ab} \pm \ems W_{ab} } = - \Om^{(\mp)} _{\upsi a b } ~,
\end{array}
\label{LorentzConnectionRules}
\end{equation}
with $h_{a b c} = e_a{}^\m e_b{}^\n e_c{}^\r\,h_{\m\n\r} = H_{a b c}$, $h_{\m\n\r}$ being the reduced field strength previously defined in \eqref{ReducedFieldStrength}. It is now very easy to know the leading order T-duality transformation of each component of the spin connection following the rules \eqref{ReducedRules}, because $\o_{a b c}, h_{a b c}$ and $e_\m{}^a$ are invariants and $V_{a b} \leftrightarrow W_{a b}$. The behavior under T-duality depends on the type of index where $\upsi$ appears and the sign of the torsion:
\begin{equation}
\begin{array}{ll}
\wt \Om^{(-)}_{A b c} = \Om^{(-)}_{A b c} ~, \qquad\qquad\qquad & \wt \Om^{(-)}_{A b \upsi} = -\Om^{(-)}_{A b \upsi} ~, \\ [1.2em]
\wt \Om^{(+)}_{a B C} = \Om^{(+)}_{a B C} ~, \qquad\qquad\qquad & \wt \Om^{(+)}_{\upsi A B } = -\Om^{(+)}_{\upsi A B} ~.
\end{array}
\label{TorsionfulConnectionRulesSimplified}
\end{equation}
Thereby we see that for $\Om^{(-)}_{A B C}$ the relevant indices for the T-duality parity sign are the last two, while only the first matters for $\Om^{(+)}_{A B C}$.

Now that we have explained how to compute $\wt \Om^{(\pm)}_{AB}{}^C$, we present the complete $(a,b)$-corrected T-duality transformation rules for $G_{M N}$, $B_{M N}$ and $\Phi$. For the latter two fields, $\wh B_{M N}$ and $\wh \Phi$ read:
\begin{equation}
\begin{array}{l}
\widehat{B}_{\m\n} = \wt B_{\m \n} + \displaystyle\sum_{k=\pm} \displaystyle\frac{a_k}{4} \; \frac{2}{\gyy} \prt{\Om^{(k)\,2}_{\psi[\m }-\frac{\Om^{(k)\,2}_{\psi\psi}}{\gyy} G_{\psi[\m }} B_{\psi \n]} ~, \\ [1.7em]
\widehat{B}_{\psi \m } = \wt B_{\psi \m} + \displaystyle\sum_{k=\pm} \displaystyle\frac{a_k}{4} \; \frac{1}{\gyy} \prt{\Om^{(k)\,2}_{\psi \m} -\frac{\Om^{(k)\,2}_{\psi\psi}}{\gyy} G_{\psi\m}} ~,
\end{array}
\label{GeneralCorrectedRulesB}
\end{equation}
and
\begin{equation}
e^{-2\wh \Phi}\sqrt{-\wh G} = e^{-2\Phi} \sqrt{-G} ~,
\label{e2PhiG}
\end{equation}
where $\Om^{(k)\,2}_{M N}$ is defined as
\begin{equation}
\Om^{(k)\,2}_{MN} :=  \Om^{(k)}_{M A}{}^B\, \Om^{(k)}_{N B}{}^ A ~;
\label{o2def}
\end{equation}
a similar definition holds for $\wt{\Om}^{(k)\,2}_{MN}$ in terms of $\wt{\Om}^{(k)}_{M A}{}^B$. Regarding the metric, we remind the reader that due to the lack of Lorentz invariance $E_M{}^A$ ---and not $G_{M N}$---, together with $B_{M N}$ and $\Phi$, provide the actual degrees of freedom of the generalized Bergshoeff-de Roo action \eqref{BergshoeffdeRoo}. Nevertheless, we shall not give explicit rules for $\wh E_M{}^A$ at first; instead, we present formulas for the $(a,b)$-corrected T-duality transformed field $\wh G_{M N}$ and thereafter explain how to obtain $\wh E_M{}^A$ from $\wh G_{M N}$. Our results read:\footnote{The rules are valid when the initial vielbein is of the form \eqref{KMVielbein}, which is the one we will use throughout this paper. See appendix \ref{AppendixOtherVielbeins} for the derivation and a possible extension to a general vielbein.}
\begin{eqnarray}
\Gh_{\mu \n} \eq \wt G_{\mu \n} + \sum_{k=\pm} \frac{a_k}{4}\prt{ \wt {\Om}^{(k)\,2}_{\m \n} -\Om^{(k)\,2}_{\m\n} + \frac{ 2\Om^{(k)\,2}_{\psi(\m} G_{\n)\psi}}{\gyy} - \frac{ \Om^{(k)\,2}_{\psi\psi}}{\gyy^2} \prt{G_{\psi \m }G_{\psi \n} - B_{\psi \m }B_{\psi \n}}} ~, \nn \\ [0.5em]
\Gh_{\psi \m } \eq  \wt G_{\psi \m }  + \sum_{k=\pm} \frac{a_k}{4}\prt{ \wt{\Om}^{(k)\,2}_{\psi \m } -\frac{\Om^{(k)\,2}_{\psi\psi} B_{\psi \m }}{\gyy^2}} ~, \label{GeneralCorrectedRulesMetric} \\ [0.5em]
\Gh_{\psi\psi} \eq \wt G_{\psi\psi}   + \sum_{k=\pm} \frac{a_k}{4}\prt{\wt{\Om}^{(k)\,2}_{\psi\psi} + \frac{\Om^{(k)~2}_{\psi\psi}}{\gyy^2}} ~, \nn
\end{eqnarray}
Notice that the $\wt \Om^{(k)\,2}_{M N}$ are always multiplied by $a_k$ in (\ref{GeneralCorrectedRulesMetric}). For this reason, when applying the rules above it is enough to compute these quantities to leading order, as the sub-leading term becomes irrelevant when multiplied by $a_k$.

Once $\wh G_{M N}$ is calculated using \eqref{GeneralCorrectedRulesMetric}, one needs a vielbein $\wh E_M{}^A$ that along with $\wh B_{M N}$ and $\wh \Phi$ solve the equations of motion \eqref{BREoM}. The dual vielbein $\wh E_M{}^A$, as any other quantity, can be written as the sum of a leading order part plus terms linear in $a_k$:
\begin{equation}
\wh E_M{}^A = (\wh E_M{}^A)^{(0)} + (\wh E_M{}^A)^{(1)} ~.
\end{equation}
The equations of motion are not local Lorentz invariant, so not any vielbein of $\wh G_{M N}$ will solve them; it is necessary and sufficient that $(\wh E_M{}^A)^{(0)} = (\wt E_M{}^A)^{(0)}$. All vielbeins of $\wh G_{M N}$ which differ only in $(\wh  E_M{}^A)^{(1)}$ are related by Lorentz transformations of the form $\delta_A{}^B + \cO(a,b) \Lm'_A{}^B$, which are actually symmetries of the equations of motion to the order we are working. This property follows easily because the only parts in the action which are not Lorentz covariant are the Chern-Simons terms appearing in $H'$ (\ref{barh}), but the compensating modification of $B_{M N}$ will be $\cO(a,b)^2$ and therefore negligible. We arrive at the same conclusion from the anomalous Lorentz transformations \eqref{AnomalousLorentzTransformation} when $\Lm = \cO(a,b) \Lm'$.

\section{Entropy considerations}
\label{SectionEntropy}

Let us derive the entropy formula for a solution of the theory (\ref{BergshoeffdeRoo}) exhibiting a bifurcate Killing horizon and discuss its behavior under anomalous Lorentz transformations.

\subsection{Generalized Wald procedure: theoretical introduction} \label{TheoreticalIntroductionEntropy}

Let us introduce in full generality the method that we will employ to derive the entropy formula. We shall follow the conventions and line of reasoning of \cite{JacobsonMohd}. It is very important to guarantee that the entropy satisfies the first law of black hole thermodynamics; the subtleties concerning this requirement were analyzed in \cite{Tachikawa} and \cite{Compere:2006my}.

Our starting point is a Lagrangian $(D+1)$-form $L = \e\,\mathcal{L} $ (with $\e$ the volume form) which, under a general variation, satisfies:
\begin{equation}
\delta L = E_i\,\delta \Psi^i + \dd \theta (\Psi, \d \Psi) ~,
\end{equation}
where $\Psi = \{\Psi^i\}$ stands for all of our fundamental fields, $E_i = 0$ are the equations of motion and the second term is a total derivative. A symplectic current can be defined as:
\begin{equation}
\Omega(\Psi, \d_1\Psi, \d_2\Psi) = \d_1 \theta(\Psi, \d_2\Psi) - \d_2 \theta(\Psi, \d_1\Psi) ~,
\label{SymplecticCurrent}
\end{equation}
where $\delta_1$ and $\delta_2$ are two generic and independent infinitesimal variations. This quantity will be relevant in deriving an explicit form of the first law. For the moment, let us consider generalized variations of the fields $\d_{\Gamma} \Psi$, where $\Gamma$ represents the set of parameters of the transformation containing at least a vector field $\zeta$ corresponding to diffeomorphisms.\footnote{$\Gamma$ might contain extra parameters which account for other symmetries of the theory as well. For the action \eqref{BergshoeffdeRoo}, we will have $\Gamma = (\zeta, \lambda, \beta)$, where $\lambda$ and $\beta$ are the parameters of the anomalous Lorentz and gauge transformations, respectively.} This variation is a generalized version of the Lie derivative and it must be a symmetry of our theory, in the sense that
\begin{equation}
\d_{\Gamma} L = \mathcal{L}_\zeta L + \dd \Xi_{\Gamma} = \dd \left( i_\zeta L + \Xi_{\Gamma} \right) ~.
\end{equation}
This allows us to define the Noether current \cite{Iyer:1994ys}
\begin{equation}
j_{\Gamma} = \theta(\Psi, \d_{\Gamma} \Psi) - i_\zeta L - \Xi_{\Gamma} ~,
\label{jGamma}
\end{equation}
whose divergence vanishes on-shell, $\dd j_{\Gamma} \cong 0$ ($\cong$ stands for equality on-shell), thereby
\begin{equation}
j_{\Gamma} \cong \dd Q_{\Gamma} ~.
\label{divjGamma}
\end{equation}
This defines the charge $Q_{\Gamma} = Q_{\Gamma}(\Psi)$. We need to study now the transformation law of $\theta$ in order to obtain the first law of thermodynamics. In general, we write $\delta_{\Gamma} \theta(\Psi, \d \Psi)$ in the following form
\begin{equation}
\delta_{\Gamma} \theta(\Psi, \d \Psi) = \mathcal{L}_\zeta \theta(\Psi, \d \Psi) + \Pi_{\Gamma} (\Psi, \d \Psi) ~,
\end{equation}
where $\Pi_{\Gamma} (\Psi, \d \Psi)$ accounts for the non-covariant part ---{\it i.e.}, not captured by the Lie derivative--- of $\theta$ \cite{Tachikawa}. Calculating $\d \d_{\Gamma} L$ in two possible ways (using $\d \d_{\Gamma} = \d_{\Gamma} \d$) we obtain $\dd \d \Xi_{\Gamma} \cong \dd \Pi_{\Gamma} (\Psi, \d \Psi)$, thereby:
\begin{equation}
\dd \Sigma_{\Gamma} (\Psi, \d \Psi) \cong \Pi_{\Gamma} (\Psi, \d \Psi) - \d \Xi_{\Gamma} ~.
\end{equation}
Finally, applying $\delta$ to (\ref{jGamma}) ---and after some algebra--- we can demonstrate that the symplectic current evaluated on-shell reads
\begin{equation}
\Omega (\Psi, \d\Psi, \d_{\Gamma} \Psi) \cong \dd \left[ \d Q_{\Gamma} - i_{\zeta} \theta(\Psi, \d\Psi) - \Sigma_{\Gamma} (\Psi, \d\Psi) \right] ~.
\end{equation}
Defining $k_{\Gamma} (\Psi, \delta \Psi) := \delta Q_{\Gamma} - i_\zeta \theta (\Psi, \d\Psi) - \Sigma_{\Gamma}(\Psi, \d\Psi)$, where in the first term we are only varying the fields of our theory (and not the parameters $\Gamma$), we have that
\begin{equation}
\Omega (\Psi, \d\Psi, \d_{\Gamma} \Psi) \cong \dd k_{\Gamma} \left( \Psi, \d\Psi \right) ~.
\label{SymplecticCurrentCharge}
\end{equation}
This can be understood as a conservation law for the charge $k_{\Gamma} (\Psi, \d \Psi)$ between two infinitesimally close field configurations provided that $\dd k_{\Gamma} (\Psi, \d \Psi) \cong 0$. In order to guarantee this, we will restrict ourselves to symmetry transformations which vanish on-shell, $\d_\G \Psi \cong 0$, since being $\Omega (\Psi, \d\Psi, \d_{\Gamma} \Psi)$ bilinear in the variations this makes the left hand side of the previous equation equal to zero.

Let us concentrate then on a particular set of symmetry transformations of our action \eqref{BergshoeffdeRoo} that generate the entropy charge when they vanish on a particular solution:\footnote{Note that we have a change of sign with respect to \cite{JacobsonMohd} in the definition of $\lambdaE^{AB}$, due to the different conventions used for Lorentz transformations. This implies $\lambdaE_{AB} \inb = - \kappa\,n_{AB}$. }
\begin{equation}
\begin{array}{lcl}
\dLLx E^A &=& \LieD E^A +E^B \lambdaE_B{}^A \cong 0 ~, \quad\qquad \dLLx \Phi = \LieD \Phi \cong 0 ~, \\ [0.8em]
\dLLx B &=& \LieD B - \displaystyle \frac{a}{4} \dd\lambdaE_A{}^B \wedge \Omega^{(-)}{}_B{}^A + \frac{b}{4} \dd\lambdaE_A{}^B \wedge \Omega^{(+)}{}_B{}^A + \dd \a_{\xi}  \cong 0 ~,
\end{array}
\label{LieLorentzTransformationsExample}
\end{equation}
where
\begin{equation}
\lambdaE^{AB} := \LieD (E^{[A})_S (E^{B]})^S ~,
\label{lambdaEAB}
\end{equation}
$\xi$ is the Killing field generating the horizon and $\a_\xi$ is a suitable gauge parameter ensuring $\d_\xi B \cong 0$ (we will discuss this choice later on). The transformations $\d_\xi$ denote exactly the same thing as $\d_\G$ for $\Gamma=(\xi,\lxieab,\a_\xi)$. Furthermore, notice that $\xi$ vanishes at the bifurcation surface because we assume a bifurcate Killing horizon.\footnote{This makes the entropy computations easier in specific cases. In general, terms of $k_{\xi}$ that are linear in $\xi$ will not contribute when evaluated at the bifurcation surface. For this reason, the relevant terms must have at least one derivative. In the following sections this will be made more precise.} 

Integrating then \eqref{SymplecticCurrentCharge} on a hypersurface with boundaries at $\mathcal{B}$ and at infinity we obtain:
\begin{equation}
\int_{\mathcal{B}} k_{\xi} (\Psi, \d \Psi) \cong \int_{\infty} k_{\xi} (\Psi, \d \Psi) ~.
\end{equation}
This is the fundamental result behind the first law of thermodynamics. We will not be concerned here with the form of the right hand side term, which should contain the variation of all the charges (energy, angular momentum, gauge charges, \dots) assuming that the fields are regular at the bifurcation surface $\mathcal{B}$. However, the left hand side is $T_H \d S$ allowing us to define the temperature and entropy as
\begin{equation}
T_H = \frac{\kappa}{2 \pi} ~, \qquad \d S = \frac{2 \pi}{\kappa} \int_{\mathcal{B}} k_{\xi} (\Psi, \d \Psi) \Big|_{\xi \to 0,  \nabla_M \xi_N \to \kappa\,n_{MN}} ~,
\end{equation}
where we have employed $\xi \inb = 0$ and $\nabla_M \xi_N \inb = \kappa\,n_{MN}$, provided that $\xi$ is properly normalized and $n_{MN}$ is the binormal to $\mathcal{B}$. 
The variation $\d S$ can be written in a different form under some extra assumptions. First of all, $\xi$ vanishes at the bifurcation surface, so the term $i_\xi \theta (\Psi, \d \Psi)$ does not contribute to the integral in $\mathcal{B}$ if our fields are all regular. In this paper, we will work with exactly invariant lagrangians ($\Xi_{\Gamma} = 0$) and our $\theta(\Psi, \d \Psi)$ will also be taken such that $\Sigma_{\xi} (\Psi, \d \Psi)$ has no relevant contribution at the bifurcation surface on-shell. We are then left with:
\begin{equation}
\d S = \frac{2 \pi}{\kappa} \,\d\!\!\int_{\mathcal{B}} Q_{\xi} (\Psi) \Big|_{\xi \to 0, \nabla_M \xi_N \to \kappa\,n_{MN}} ~,
\end{equation}
where $Q_{\xi}$ was introduced in (\ref{divjGamma}). Finally, since terms linear in $\xi$ in the integral will not contribute at the bifurcation surface, we find that the relevant contribution in $Q_{\xi} (\Psi)$ is linear in $\nabla_M \xi_N$, and thus linear in $\kappa$ when evaluated at $\mathcal{B}$. The surface gravity $\kappa$ is constant (zeroth law), and $\d \kappa = 0$, understanding $\d$ as a variation leaving the Killing field $\xi$ fixed \cite{Wald1993,Iyer:1994ys}. As a consequence, under the previous assumptions we obtain an expression for the entropy as an integral over the bifurcation surface:
\begin{equation}
S = 2 \pi \int_{\mathcal{B}} Q_{\xi} (\Psi) \Big|_{\xi \to 0, \nabla_M \xi_N \to n_{MN}} ~.
\label{EntropyFormula}
\end{equation}
In the next section we will present the computation of the entropy charge for the generalized Bergshoeff-de Roo action \eqref{BergshoeffdeRoo}. 

\subsection{Entropy of the generalized Bergshoeff-de Roo action}
\label{EntropyBdR}

Let us now apply the previous general argument to the generalized Bergshoeff-de Roo action \eqref{BergshoeffdeRoo}. For the sake of simplicity, we will work in this section (and only here) with the Killing field normalized so that $\nabla_M \xi_N \inb = n_{MN}$. In addition, we will split the action as $\mathcal{I}_{\rm BdR} = \mathcal{I}_0 + \mathcal{I}_{H'^2} + \mathcal{I}_{R^2}$, where
\begin{equation}
\begin{array}{lcl}
\mathcal{I}_0 \eq \displaystyle\int \epsilon ~ e^{-2 \Phi} \left[ R - 2 \Lambda + 4 \nabla_M \Phi \nabla^M \Phi \right] ~, \\ [1.2em]
\mathcal{I}_{H'^2} \eq -\displaystyle\frac{1}{12} \int \epsilon ~ e^{-2 \Phi} H'_{MNR} H'^{MNR} = - \frac{1}{2} \int e^{-2 \Phi}  \star H'\wedge H' ~, \\ [1.2em]
\mathcal{I}_{R^2} \eq \displaystyle\sum_{k=\pm} \frac{a_k}{8} \int \epsilon ~ e^{-2 \Phi} R^{(k)}_{MNA}{}^{B} R^{(k)}{}^{MN}{}_B{}^{A} = \sum_{k=\pm} \frac{a_k}{4} \int e^{-2 \Phi} \star R^{(k)}_A{}^B \wedge R^{(k)}_B{}^A ~.
\end{array}
\label{Action0}
\end{equation}
We have performed an innocuous integration by parts in $\mathcal{I}_0$ in order to obtain a more convenient form of the dilaton  kinetic term. Given this action, we have to follow the general lines we presented in the previous section, starting from the computation of the boundary term $\theta (\Psi, \d \Psi)$ and going all the way to the final result for the entropy charge $Q_\xi (\Psi)$. In Appendix \ref{AppendixEntropyComputation} we show in full detail how this is achieved. Here we will only quote the main results. Taking into account that we are using \eqref{LieLorentzTransformationsExample} as the symmetry transformations to compute the entropy charge, we obtain:\footnote{The reason for this splitting is made clear in Appendix \ref{AppendixEntropyComputation}.}
$$
Q_{\xi} = Q_0 + Q_{H'^2} + Q_{R^2} + Q_{\alpha_{\xi}} ~,
$$
where
\begin{equation} \label{Charge0BdR}
Q_{0} = - 2 e^{-2 \Phi} \nabla^{M} \xi^{N} \left( \dd^{D-1} x \right)_{MN} + \ldots ~,
\end{equation}
is the contribution coming from $\mathcal{I}_0$, whereas
\begin{equation}
\begin{array}{ccl}
Q_{H'^2} \eq e^{-2 \Phi}  \star H \wedge \left[ 2 \gm \Omega_{A}{}^{B} + \gp \cH_{A}{}^{B} \right] \lambdaE_{B}{}^{A} + \ldots ~, \\ [0.8em]
Q_{R^2} \eq - e^{-2 \Phi} \bigg[ 2 \gp \star \left( R_A{}^B + \frac{1}{4} \cH_A{}^C \wedge \cH_C{}^B \right) \\
& & \qquad\qquad\qquad + \,\gm \star \left( \dd \cH_A{}^B + 2 \Omega_A{}^C \wedge \cH_C{}^B \right) \bigg] \lambdaE_B{}^A + \dots ~,
\end{array}
\label{ChargesHRBdR}
\end{equation}
with $\cH_{A}{}^{B}$ given by \eqref{H1form} and
\begin{equation}
\gamma_\pm = \mp \frac{a \pm b}{4} ~,
\label{ParametersGamma}
\end{equation}
are the contributions of, respectively, $\mathcal{I}_{H'^2}$ and $\mathcal{I}_{R^2}$, except those arising from the gauge transformation parameter $\a_{\xi}$ in \eqref{LieLorentzTransformationsExample}. This latter contribution is isolated in $Q_{\alpha_{\xi}}$,
\begin{equation}
Q_{\alpha_{\xi}} = 6 \mathbb{E}^{MNR} (\a_\xi)_R \left( \dd^{D-1} x \right)_{MN} ~,
\label{ChargeBBdR}
\end{equation}
where $\mathbb{E}^{MNR} := T^{[MNR]} - \nabla_Q S^{Q[MNR]}$, with
\begin{equation}
T^{MNR} = \frac{\partial \mathcal{L}}{\partial H_{MNR}}~, \qquad S^{QMNR} = \frac{\partial \mathcal{L}}{\partial \nabla_Q H_{MNR}} ~,
\label{DefinitionsTandS}
\end{equation}
where $L = \e\,\mathcal{L}$ is our full Lagrangian. Notice that the equation of motion for the $B$-field is simply $\nabla_M \mathbb{E}^{MNR} \cong 0$ (see Appendix \ref{AppendixEntropyComputation} for details).

Some comments are in order here. First of all, when convenient, we are using a notation for differential forms that follows \cite{Compere:2006my} and is presented in appendix \ref{AppendixEntropyComputation}. The dots in the entropy charges denote omitted terms which do not contribute when evaluated at the bifurcation surface (that is, terms proportional to $\xi^M$ thereby vanishing from the assumption of regularity applied to all fields). Finally, it is important to remember that $\alpha_\xi$ is not a free parameter of a gauge transformation. It is determined (up to the addition of a closed form) from the condition that the variation of the $B$ field given by \eqref{LieLorentzTransformationsExample} has to vanish on-shell. In Section 4 we will set $\alpha_\xi = 0$ in a region near the horizon but, for the moment, let us keep track of $\alpha_\xi$ as it will be necessary to show the invariance of the entropy under anomalous Lorentz transformations.

Before presenting the full form of the entropy, it is illustrative to see which would be its value if we had only considered the action $\mathcal{I}_0$. It would have been given by
\begin{equation}
\begin{array}{lcl}
S_0 &=& 2 \pi \displaystyle\int_{\mathcal{B}} Q_{0,\xi} (\Psi) \Big|_{\xi \to 0, \nabla_M \xi_N \to n_{MN}} \\ [1.2em]
&=& -4 \pi \displaystyle\int_{\mathcal{B}} e^{-2 \Phi} n^{MN} \left( \dd^{D-1} x \right)_{MN} = 4 \pi \int_{\mathcal{B}} e^{-2 \Phi} \bar{\e} ~,
\end{array}
\end{equation}
where we used $\left. \left( \dd^{D-1} x \right)_{MN} \right|_{\mathcal{B}} = n_{MN} \,\bar{\e}/2$ and $n_{M N} n^{M N} = -2$, $\bar{\e}$  being the induced volume form on the bifurcation surface $\mathcal{B}$ (see Appendix \ref{AppendixEntropyComputation}). This is just the expected Einstein-Hilbert contribution corrected by the dilaton term. Now, the entropies coming from the other terms can be obtained after some manipulations and are given by
\begin{eqnarray}
S_{H'^2} \eq 4 \pi \int_{\mathcal{B}} e^{-2 \Phi} \,\star H \wedge \left( \gm \Omega^{AB} + \frac{\gp}{2} \cH^{AB} \right) n_{AB} ~, \label{entropyBR1} \\ [0.7em]
S_{R^2} \eq - 4 \pi \gp \int_{\mathcal{B}} e^{-2 \Phi} \,\star\left( R^{AB} + \frac{1}{4} \cH^{AC} \wedge \cH_C{}^{B} \right) n_{AB} ~, \label{entropyBR2} \\ [0.7em]
S_{\alpha_\xi} \eq 6 \pi \int_{\mathcal{B}} \bar{\e}\, \mathbb{E}^{M N R} n_{MN} (\a_\xi)_R ~.
\label{entropyBR3}
\end{eqnarray}
All in all, writing the fields in tensorial form and using the fact that, given that the binormal can always be written as $n_{M N} = 2 v_{[M} w_{N]}$, where $v,w$ are some 1-forms \cite{Compere:2006my}, it obeys $n_{M [N} n_{R S]} = 0$, the two terms of the form $\gp H H n n$ in \eqref{entropyBR1} and \eqref{entropyBR2} can be combined together yielding the following result for the entropy:
\begin{equation}
\begin{array}{l}
\displaystyle S = S_0 - 2\pi \int_{\mathcal{B}} \bar{\e} \, e^{-2\Phi} \bigg[ \gp \left( R^{MNRS} - \frac{3}{4} H^{TMN} H_T{}^{RS} \right) n_{MN} n_{RS} \\ [1.3em]
\qquad\qquad\qquad\qquad\qquad\qquad\qquad - \gm\, H^{TMN} \Omega_T{}^{RS} n_{MN} n_{RS} \bigg] + S_{\alpha_\xi} ~ ,
\end{array}
\label{FullEntropyBR}
\end{equation}
where $\Om_{T}{}^{RS} = \Om_T{}^{AB} E_A{}^R E_B{}^S$. In Appendix \ref{AppendixFieldRedefinition} we use the field redefinition method to derive the entropy for $\alpha_\xi = 0$.

Given that our theory is invariant under anomalous Lorentz transformations \eqref{AnomalousLorentzTransformation}, we strongly expect this symmetry to be present in the entropy as well. Let us check this explicitly by considering the following transformation to a new set of fundamental fields:
\begin{equation}
\begin{array}{lcl}
E'^A & = & E^A + E^B \Lambda_B{}^A ~, \qquad\qquad \Phi'= \Phi ~, \\ [0.8em]
B' & = & B + \gm \Omega_A{}^B \wedge \dd\Lambda_B{}^A + \displaystyle \frac{\gp}{2} \cH_A{}^B \wedge \dd\Lambda_B{}^A ~.
\end{array}
\end{equation}
The new $\lambdaEp$ for this vielbein becomes:
\begin{equation}
\lambdaEp_B{}^A = \lambdaE_B{}^A + \lambdaE_B{}^C \Lambda_C{}^A - \Lambda_B{}^C \lambdaE_C{}^A - \LieD \Lambda_B{}^A ~.
\label{TransformedLambda}
\end{equation}
Now, for these Lorentz transformed fields we must be sure that the symmetry transformations we employ to compute the entropy \eqref{LieLorentzTransformationsExample} vanish on-shell. The new transformations are related to the old ones by:
\begin{equation}
\begin{array}{lcl}
\dLLx E'^A  & = &  \dLLx E^A + (\dLLx E^B) \Lambda_B{}^A ~, \\ [0.8em]
\dLLx B' & = & \dLLx B + \gm \dLLx \Omega_A{}^B \wedge \dd\Lambda_B{}^A \\ [0.7em]
& & \quad + \displaystyle \frac{\gp}{2} \dLLx \cH_A{}^B \wedge \dd\Lambda_B{}^A + \dd\left[ \d_\Lambda \alpha'_{\xi} - 2 \gm \lambdaE_A{}^B  \dd\Lambda_B{}^A \right] ~,
\end{array}
\label{TransformedLieLorentzTransformation}
\end{equation}
where $\d_\Lambda \a_\xi = \a'_{\xi} - \a_\xi$. It follows from $\d_\xi \Psi=0$ that $\d_\xi \Om_A{}^B=0$ and $\d_\xi \cH_A{}^B=0$. Consequently, we need $\dd[\d_\Lambda \a_\xi -  2 \gm \lambdaE_A{}^B  \dd\Lambda_B{}^A] =0$ in order to satisfy $\d_\xi B' = 0$. We choose $\d_\Lambda \a_\xi =  2 \gm \lambdaE_A{}^B  \dd\Lambda_B{}^A$; that is, the choice of the suitable gauge parameter $\a_\xi$ must generically be changed under anomalous Lorentz transformation in order to guarantee $\d_\xi B'=0$.

Let us go back to our computation of the anomalous Lorentz invariance of the entropy. Given the fact that, to first order, the only non-Lorentz covariant terms in the entropy \eqref{FullEntropyBR} are $S_{\alpha_\xi}$ and the one containing the spin-connection, we conclude that:
\begin{equation}
\begin{array}{ccl}
\d_\Lambda S & = & 2 \pi \displaystyle \int_{\mathcal{B}} \bar{\e} \, e^{-2 \Phi} \left[ \gm H^{MNR} \d_\Lambda \Omega_R{}^{AB} n_{AB} - \frac{1}{2} H^{MNR} \d_\Lambda (\a_\xi)_R \right] n_{MN} \\ [1.2em]
& = & 2 \pi \gm \displaystyle \int_{\mathcal{B}} \bar{\e} \, e^{-2 \Phi}  H^{M N R} \partial_R \Lambda^{AB} \left[ n_{AB} + \lambdaE_{AB} \right]n_{M N} = 0 ~, 
\end{array}
\label{LorentzInvarianceEntropy}
\end{equation}
where we have used $\lambdaE_{AB} \inb = - n_{AB}$, the fact that $S^{QMNR} = \mathcal{O}(\g_{\pm})$, and
\begin{equation}
T^{MNR} \approx - \frac{1}{6} e^{-2 \Phi} H^{MNR}  ~ ,
\end{equation}
where $\approx$ denotes that the quantities differ at most at linear order in $a,b$. This shows that the entropy is invariant under infinitesimal anomalous Lorentz transformations around a generic vielbein. Note the key role played by the parameter $\a_\xi$. To get an invariant entropy, one needs to impose an invariant stationarity condition like $\d_\xi B = 0$, and this is only possible by means of $\a_\xi$ and its non-trivial anomalous Lorentz `transformation'. We can summarize by saying that stationarity and anomalous Lorentz invariance are compatible via gauge symmetry.

\section{T-duality invariance of the entropy and temperature}
\label{SectionInvariance}

In this section we show that the entropy is exactly invariant to linear order in $a,b$ under the corrected T-duality rules. The invariance occurs for all values of $a$ and $b$, even those not corresponding to effective string theories. Furthermore, the horizon temperature turns out to be invariant as well.

\subsection{Convenient coordinates and vielbein near the horizon}
\label{SubsectionCoordinatesAndVielbein}

We will deal with horizons of the kind described in \cite{Racz:1992bp}. Their main characteristic is that they are stationary spacetimes with a bifurcate Killing horizon. Every regular Killing horizon with constant surface gravity $ \k \neq 0 $ is of bifurcate type and viceversa; we can take $\kappa > 0$ without loss of generality. These horizons can be extended to include a regular bifurcation surface $\mathcal B$,\footnote{$\mathcal B$ can be defined as the locus of vanishing null Killing vector $\xi$; see \cite{Horowitz:1993wt}.} where we will evaluate the entropy. It is very convenient to use a generalization of the Kruskal coordinates in some neighborhood of the horizon. As in the Schwarzschild black hole, they cover smoothly an entire neighborhood of the horizon, and in particular the bifurcation surface. The general line element in any spacetime dimension reads
\begin{equation}
\dd s^2 = G\,\dd U \dd V + V F_{\a'}\,\dd U\dd x^{\a'} + \g_{\a'\b'}\,\dd x^{\a'} \dd x^{\b'} ~,
\label{GeneralKruskal}
\end{equation}
where $G, F_{\a'}$ and $\g_{\a' \b'}$ are regular functions. The null Killing field in these coordinates is given by $\xi = \k (U \partial_U - V \partial_V)$, where $\k$ is the surface gravity with respect to $\xi$.\footnote{In asymptotically flat spacetimes, it is customary to normalize the Killing vector such that $\xi^2 = -1$ at infinity. But this criterion cannot be applied in all cases, for example in AdS spacetimes. Therefore we will not impose any particular normalization.} The coordinates labeled with a primed Greek index, $x^{\a^\prime}$, include $x^\a$, $\a = 1, \ldots, D-2$, and $\psi$, a coordinate adapted to the U(1) symmetry required for T-duality, as in Section \ref{SectionCorrectedRules}. Consequently $\partial_\psi G_{M N}=0$ and the same holds for $G,F_{\a'},\g_{\a' \b'}$. Now we choose a vielbein for \eqref{GeneralKruskal} as\footnote{To identify this vielbein one should first find $E^{\upsi}$ as defined in \eqref{VierbeinKruskal}. After that, we take a null vielbein of the form $\dd s^2 = -2 E^+ E^- + (E^2)^2 + (E^3)^2 + \cdots + (E^{\underline{\psi}})^2$ with $E^+ \propto \dd U$, and convert it to a usual vielbein.}
\begin{eqnarray}
E^0 \eq \frac{1}{\sqrt{2}} \bigg[ \prt{\half + \frac{1}{4} V^2 \emds F_\psi^2}\,\dd U - G\,\dd V + V(\emds F_\psi \g_{\psi\a}-F_\a)\,\dd x^\a \bigg] ~, \nn \\ [0.5em]
E^1 \eq \frac{1}{\sqrt{2}} \bigg[ \prt{\half - \frac{1}{4} V^2 \emds F_\psi^2}\,\dd U + G\,\dd V - V(\emds F_\psi \g_{\psi\a}-F_\a)\,\dd x^\a \bigg] ~,\nn \\ [0.6em]
E^i \eq \dd x^\a e_\a{}^i  ~, \quad i = 2, \ldots, D-1 ~, \nn \\ [0.6em]
E^{\underline{\psi}} \eq \frac{1}{2} V \ems F_\psi \dd U + e^{-\s} \g_{\psi \a} \dd x^\a + e^\s \dd \psi ~,
\label{VierbeinKruskal}
\end{eqnarray}
where the $e_\a{}^i$ constitute a vielbein for $\g_{\a \b}$; {\it i.e.}, $\delta_{ij} e^i e^j = \g_{\a\b}\,\dd x^\a \dd x^\b$. This vielbein choice is convenient for three reasons. The first is that it contains $\dd \psi$ only in the component $E^{\upsi}$ and therefore it is of the form \eqref{KMVielbein}; consequently, the corresponding uncorrected simple rules \eqref{ReducedRules} and \eqref{LorentzConnectionRules} apply to it. The second is that all components are smooth and so is the inverse vielbein, $E_A{}^M$. Therefore, the connection components $\Om_{M A}{}^B$ are regular as well, even on the bifurcation surface $\bif$. Notice that regularity is crucial in the derivation of the entropy formula \cite{JacobsonMohd} performed in Section \ref{SectionEntropy}. Note also that in this vielbein the stationarity condition $\d_\xi E_M{}^A = \li E_M{}^A + E_M{}^B \lxieba = 0$ is fulfilled with $\lxieba$ being the generator of a uniform boost along the $E^1$ direction. This is a consequence of:
\begin{equation}
\li E_M{}^0 = \k E_M{}^1 ~, \qquad \li E_M{}^1 = \k E_M{}^0 ~,
\label{LieVierbeinKruskal}
\end{equation}
while $\li E_M{}^i = \li E_M{}^{\underline{\psi}} = 0$. Using \eqref{LieLorentzTransformationsExample}, we see that $(\l_\xi^E)_{01} = - (\l_\xi^E)_{10} = -\k$ while the remaining components vanish; therefore we have
\begin{equation}
\dd \lxieab = 0 ~.
\label{dlambdavanishes}
\end{equation}
We will consider a two-form field $B$ such that $\li B = 0$. This leads to the third good feature of the vielbein: the stationarity condition \eqref{LieLorentzTransformationsExample} for $B$ is simplified with \eqref{dlambdavanishes} to the form $\d_\xi B = \dd \a_\xi = 0 $. In this way we will take $\a_\xi=0$ in what follows. The reader should keep in mind that $\dd \lxieba = 0$ only holds in a neighborhood of the horizon covered by $U, V, x^{\a'}$. The knowledge of the fields in such neighborhood is the only necessary data to compute the entropy and the temperature, which are invariant under anomalous Lorentz transformations. Furthermore, we demand also the stationarity condition on the dilaton $\d_\xi \Phi = \li \Phi = 0 $. Finally, the vielbein also fulfills $\partial_\psi E_M{}^A = 0$ while for the matter fields we require the U(1) symmetry for T-duality:
\begin{equation}
\partial_\psi B_{M N} = 0 ~, \qquad \partial_\psi \Phi = 0 ~.
\end{equation}
We will also demand $G_{\psi \psi} \neq 0 $ everywhere to prevent curvature singularities in the T-dual solution.

\subsection{Invariance of the entropy and temperature} 
\label{SectionEntropyInvariance}

In this subsection we compare the horizon entropies before and after T-duality. As a matter of fact, it turns out that they are the same for all values of $a$ and $b$. This generalizes the result of the uncorrected $a = b = 0$ case \cite{Horowitz:1993wt}.

Before proceeding to compute the entropy of the T-dual solution, it is necessary to show that we actually have a bifurcate Killing horizon after the corrected T-duality rules are applied to a black hole spacetime. A basic requirement is the regularity of the dual metric, which follows from $G_{\psi \psi} \neq 0$ and the non-singular $\Omega_{M A}{}^B$ and $\Omega^{(\pm)}_{M A}{}^B$ before duality. Furthermore, in \eqref{TorsionfulConnectionRulesSimplified} we see that $\wt \Om^{(\pm)}_{M A}{}^B$ must be regular. Then, we obtain a regular dual metric when we apply the corrected T-duality rules \eqref{GeneralCorrectedRulesMetric}.

In order to have a bifurcate Killing horizon one needs a Killing vector that is null on the horizon and vanishes on a codimension-2 surface. In fact, the same Killing field $\xi$ of $G_{M N}$ will also satisfy such conditions with $\wh G_{M N}$ as the metric; in Appendix \ref{AppendixStationarityOfTheDual} we establish that $\xi$ is a Killing vector of $\wh G_{M N}$. Furthermore, $\xi^M$ does not depend on the fields and then it is the same after T-duality, vanishing on $U = V = 0$. To show that it is null and orthogonal to the horizon we follow an argument similar to that of \cite{Horowitz:1993wt}. As $\wh G_{MN}$ is regular and $\xi^M \inb = 0$, the scalars $ \xi^M \xi ^N \wh G_{MN} \inb = \xi^M (\partial_{\a'})^N \wh G_{MN} \inb = 0$. Moreover, $\li (\partial_{\a'})^M=0$, so these scalars are symmetric under $\xi$. From any point of $UV = 0$, one can get arbitrarily close to $\bif$ through the flow of the Killing vector $\xi$. By continuity, the scalars $\xi^M \xi ^N \wh G_{MN}$ and $\xi^M (\partial_{\a'})^N \wh G_{M N}$ also vanish for any point in $UV=0$. There is a spacelike codimension-2 surface where $\xi$ vanishes, namely $U = V = 0$. In the remaining points of $UV=0$, there is a non-zero normal Killing vector $\xi$ with respect to the metric $\wh G_{M N}$. Consequently, there is a bifurcate Killing horizon in $UV=0$ after the duality \cite{Horowitz:1993wt,Racz:1992bp}.

It is important to mention that the dual fields satisfy the stationarity conditions \eqref{LieLorentzTransformationsExample} with $\wh \a_\xi=0$; this is detailed in Appendix \ref{AppendixStationarityOfTheDual}. There is an aspect of the T-dual configuration which is not determined by the corrected T-duality rules; namely, the range of the dual coordinate $\wh{\Delta \psi}$. In the case of string theory, calculations using the path-integral of the underlying worldsheet description show that the ranges should be equal, $\Delta \psi = \wh{\Delta \psi}$.\footnote{For isometries corresponding to a compact U(1), $\Delta \psi = \wh{\Delta \psi} = 2\pi$; in general, $\Delta \psi$ and $\wh{\Delta \psi}$ must be reciprocal to each other \cite{Rocek:1991ps}.} We assume this is the case for all values of $a$ and $b$; otherwise, we would both spoil the entropy invariance for $a=b=0$ already found in \cite{Horowitz:1993wt} and the invariance of the action under T-duality, as long as the Lagrangian itself is invariant.

Now, we are going to present the expression for the entropy in terms of our vielbein. Before applying T-duality, $n = E^0 \wg E^1$ and $\a_\xi = 0$, which implies that the entropy is given by:
\begin{equation}
S = 2\pi\!\int_{\bif} \!\dd^{D-1}x\, e^{-2\Phi} \sqrt{G_h} \,\bigg[ 2 + 4 \gp \prt{R^{01}_{\;\;\; 01} - \frac{3}{4} H^{A01} H_{A01}} - 4 \gm \Om^{A01} H_{A01} \bigg] ~,
\label{EvaluatedEntropy}
\end{equation}
where $\dd^{D-1}x := d\psi\,\dd^{D-2}x$. After T-duality, the components of the binormal at the bifurcation surface, $n_{M N}\inb$, are the same to leading order. Actually, it is exactly the same, as explained in Appendix \ref{AppendixAreaInvariance}. In turn, this implies $\wh n_{A B}\inb \approx n_{A B}\inb$, the leading order being enough for our computation as stated before. Since $\wh \a_\xi = 0$, the integrand of the entropy after T-duality is given by the same expression \eqref{EvaluatedEntropy}, just placing a hat on each field.

The next step is to relate the integrands before and after duality. In fact, it turns out that they have both the same value. Let us elaborate on this. The factor $e^{-2\Phi} \sqrt{G_h}$ contributes to both leading and subleading order in $a$ and $b$, so we need to know how it transforms under the $(a,b)$-corrected rules \eqref{GeneralCorrectedRulesMetric}. It ends up being an invariant, and the details are presented in Appendix \ref{AppendixAreaInvariance}. It is possible to summarize the derivation by saying that the stationarity conditions \eqref{LieLorentzTransformationsExample} constrain the dual metric to the form
\begin{equation}
\wh G_{M N} \inb =
\begin{pmatrix}
0 & G_{UV}\inb & 0 \\
G_{UV}\inb & 0 & 0 \\
0 & 0 & \wh G_{\a'\b'}\inb
\end{pmatrix} ~.
\label{DualGInB}
\end{equation}
Therefore, the dual determinant factorizes when evaluated on the bifurcation surface, $\wh G\inb = \wh G_\perp\inb \,\wh G_h\inb$, where $\wh G_\perp\inb = -G_{U V}^2\inb$ is the determinant of the metric orthogonal to $\bif$. Given that $\wh G_\perp\inb = G_\perp\inb$, and noticing that the determinant before T-duality also factorizes on the bifurcation surface \eqref{VierbeinKruskal}, we obtain
\begin{equation}
e^{-2\wh \Phi} \sqrt{\wh G_h} \Big\rvert_{\mathcal{B}} = e^{-2\Phi} \sqrt{G_h}  \Big\rvert_{\mathcal{B}} ~ ,
\label{ResultsAreaInvariance}
\end{equation}
where
we relied on the invariance of $e^{-2\Phi} \sqrt{-G}$ under the general corrected T-duality rules \eqref{e2PhiG}.

We have to investigate how the expressions in the square brackets of \eqref{EvaluatedEntropy} transform under T-duality. The term contributing to the so-called {\it area law} corresponds to the first summand, which is invariant as it is constant, multiplied by $e^{-2\Phi} \sqrt{G_h}$ and integrated on the bifurcation surface. The other two summands are already $\mathcal O(a,b)$. Our vielbein \eqref{VierbeinKruskal} and its leading order T-dual are of the class specified in \eqref{KMAnsatz}, so we can use the dimensionally-reduced uncorrected T-duality rules \eqref{ReducedRules}. To do so, one has to perform first the dimensional reduction:
\begin{eqnarray}
R^{01}_{\;\;\; 01} - \frac 3 4 H^{A01}  H_{A01} &=& r^{01}_{\;\;\; 01} - \frac{3}{4} h^{a01} h_{a01} - \frac{3}{4} (e^{2\s}V^{01} V_{01} + e^{-2\s} W^{01}W_{01}) ~, \label{dimredR0101} \\ [0.7em]
\Om^{A01} H_{A01} &=& \o^{a01} h_{a01} - \half V^{01}W_{01} ~.
\label{dimredOmegaH}
\end{eqnarray}
The results for the reduction of $R_{ABCD}$ and $H_{ABC}$ were already presented in \cite{KaloperMeissner}, while we gave those of $\Om_{ABC}$ in \eqref{LorentzConnectionRules}. The reduced Riemann tensor, field strength and Lorentz-connection $r^{01}_{\;\;\; 01}$, $h^{a01}$ and $\o^{a01}$ are invariant up to $\mathcal O(a,b)$ terms, and the reduced rules \eqref{ReducedRules} imply $\wh \s \approx -\s$, $\wh V_{01} \approx W_{01}$ and $\wh W_{01} \approx V_{01}$. As a consequence, both \eqref{dimredR0101} and \eqref{dimredOmegaH} are T-dual invariant to leading order. Notice that the $V V$ and $W W$ terms stem respectively from $R$ and $H H$ in such a way that the actual relative factor $-\frac{3}{4}$ is crucial to yield the invariance. We have attained entropy invariance,
\begin{equation}
\wh S = S ~,
\label{ResultEntropyInvariance}
\end{equation}
which is the main result of our work. It is also possible to derive the T-dual invariance of the temperature, $T_H = {\k}/{2 \pi}$, using the form of $\wh G_{MN} \inb$. In fact, it is straightforward to compute the dual surface gravity $\wh \k$ in the bifurcation surface:
\begin{equation}
 \wh \k\, \wh n_M{}^N \inb  = \wh \nabla_M \xi^N \inb = \partial_M \xi^N \inb =  \k\, n_M{}^N \inb  ~.
\end{equation}
The second equality follows from $\xi^M\inb = 0$ and the latter is the consequence of $\nabla_M \xi^N \inb = \k\, n_M{}^N \inb$. Notice how $\partial_M \xi^N\inb $ does not depend on the dual fields at all. It follows from \eqref{DualGInB} that $\wh n_M{}^N \inb = n_M{}^N\inb$. In fact, the binormal is also the normalized volume form of the 2-dimensional subspace orthogonal to $\bif$. In this case the latter is spanned by $\partial_U$ and $\partial_V$, and the corresponding part of the metric does not change under the corrected T-duality rules. This means that $\wh n_M{}^N\inb = n_M{}^N\inb$, consequently:
\begin{equation} 
\wh \k = \k ~.
\label{KappaInvariance}
\end{equation}
Therefore, we have established the T-dual invariance of the temperature. This result may seem somehow expected from the fact that the corrected T-duality transformations are a sequence of field redefinitions (see \eqref{BarredFieldRedefinition} in Appendix \ref{AppendixOtherVielbeins}) followed by the uncorrected Buscher rules and, finally, corresponding inverse field redefinitions. Each of those operations are expected to preserve surface gravity on their own. Indeed, it was proven in  \cite{Jacobson:1993vj} that, in the case of a regular bifurcate Killing horizon, the surface gravity is constant irrespective of the underlying gravitational dynamics provided $G_{M N} \to G_{M N} + \Delta_{M N}$, where $\Delta_{M N}$ is a regular tensor such that $\li \Delta_{M N} = 0$. In our particular case, both conditions are satisfied for our vielbein \eqref{VierbeinKruskal} because $\Om_{M A}{}^B$ is finite and $\li \Om^{(k)\, 2}_{M N} \approx \li  \wt \Om^{(k)\, 2}_{M N} \approx 0$ (see Appendix \ref{AppendixStationarityOfTheDual} for a derivation). 

\section{Discussion and concluding remarks} 

In this work we deal with a family of perturbative four-derivative actions describing gravity coupled to a Kalb-Ramond field and a dilaton, involving two parameters $a$ and $b$ weighted by the inverse mass scale $M_\star^{-2}$ \cite{MarquesNunez}. For all values of $a$ and $b$ the Lagrangian is perturbatively invariant under T-duality. Nevertheless, only a few choices of $a$ and $b$ correspond to effective string actions, and in this sense it is possible to speak of T-duality beyond String Theory. 

Our main conclusion is that the entropy and temperature of a generic non-extremal black hole solution are invariant under T-duality, thereby extending the original analysis of \cite{Horowitz:1993wt} to next-to-leading order in the derivative expansion. This happens for all values of $a$ and $b$, whether they are stringy or not.\footnote{For a recent discussion on $\alpha^\prime$-corrected T-duality in heterotic string theory, see \cite{CanoOrtin1,ChimentoOrtin2}.} This is somehow surprising since T-duality is not expected to be a symmetry of theories based on point particles. We have therefore extended our previous results for the invariance of entropy and temperature found for the BTZ black hole well beyond its particular symmetry and dimensionality \cite{ESSV1}; it is well-known that gravity in 2+1 dimensions is special. In order to attain this result it was necessary to deal with the $(a,b)$-corrections to the T-duality rules, which we explicitly derived. This further supports the idea put forward in our previous work that T-duality may be relevant in providing physical equivalences beyond the realm of String Theory.

Another interesting result concerns the derivation of the entropy formula. In particular, its anomalous Lorentz invariance requires non-obvious stationarity conditions adapted to such symmetry. The gauge invariance of the $B$-field, while usually disregarded in most derivations of the entropy, becomes absolutely necessary in this case. It is possible to check the resulting expression with an independent derivation based on the method of field redefinitions.

The fact that bifurcate Killing horizons are mapped onto themselves with exactly the same surface gravity, generalizing the results of \cite{Horowitz:1993wt}, can be explained as follows: for a given metric, $(a,b)$-corrected T-duality is a sequence of field redefinition, uncorrected Buscher rules and another field redefinition. As mentioned in Section \ref{SectionInvariance}, each of those operations preserve surface gravity on their own. Therefore, the same must happen for their successive application.

The invariance of both the black hole entropy and temperature are in line with the generic expectations of \cite{ArvanitakisBlair} for higher derivative corrections derived from Double Field Theory with a generalized metric. Such conclusions must be taken with a grain of salt, though, since our action is derived from Double Field Theory with a generalized vielbein and a generalized anomalous Lorentz symmetry \cite{MarquesNunez}. Therefore we suspect that their conclusions do not apply exactly to our case, even if the results are consistent. It would be certainly interesting to bridge the gap between the construction presented in \cite{ArvanitakisBlair} and our results.

In the bigger picture, we would like to explore T-duality as a symmetry principle to constrain effective actions with degrees of freedom given by $G_{M N}$, $B_{M N}$ and $\Phi$. This is much more stringent than diffeomorphism invariance, and we hope that some of the appealing properties of T-duality in String Theory will be inherited by effective actions, like the equivalence of small and large compact directions or the equivalence of momentum and charge \cite{Horne:1991cn}.\footnote{For instance, it was recently suggested that T-duality might be a key ingredient solving the singularity of static, electrically neutral black hole metrics \cite{Nicolini:2019irw}.} However, there is one caveat in this program. By requiring T-duality as a symmetry principle we will certainly make it a generating technique, but it is not guaranteed that it will become a physical equivalence in the way it is in $\sigma$-models defining string theories for generic values of $a$ and $b$.

A general answer to this question for the moment seems beyond our reach; nevertheless, the analysis of the entropy and temperature in the two-parameter family of theories studied in this paper is intended as a first step. As mentioned above, we found that the invariance of entropy and temperature holds non-trivially for all $a$ and $b$; this is a necessary condition to behave as a physical duality, albeit it is unclear if it is also a sufficient condition. In fact, contrary to our expectations we could not find any distinctive behavior in the stringy cases that proves them to be special. It  therefore becomes necessary to test this proposal with further checks. The most obvious extension would be to study the behavior of the other thermodynamic quantities entering the first law; in particular, to scrutinize if the mass is invariant for asymptotically flat spacetime and momentum and charge get exchanged like in \cite{Horne:1991cn}. It would also be interesting to explore whether our results hold for higher orders in the derivative expansion. Building these actions seems a quite difficult challenge, albeit substantial progress has been achieved recently \cite{LescanoMarques,BaronLescanoMarques}.

Another important issue is to study the consistency of anomalous Lorentz transformations with corrected T-duality. Specifically, one must be sure that two solutions related by anomalous Lorentz transformation will have physically equivalent T-duals. There is good reason to think that this is the case for the bosonic action $a=b$, since it can be fully rewritten in terms of $G_{M N}$, $B_{M N}$ and $\Phi$, and the anomalous piece of the Lorentz transformation disappears \cite{MarquesNunez}. It would be interesting to see if this requirement can set stringy cases apart from generic $a$ and $b$, as the heterotic case $b=0$ also has peculiar properties under anomalous Lorentz symmetry.

\section*{Acknowledgements} 

We are grateful to Ted Jacobson, Eric Lescano, Diego Marqu\'es, Sudipta Sarkar and Yuji Tachikawa for fruitful discussions.
The work of J.D.E. is supported by the Ministry of Science grant FPA2017-84436-P, Xunta de Galicia ED431C 2017/07, FEDER, and the Mar\'\i a de Maeztu Unit of Excellence MDM-2016-0692. He wishes to thank Pontificia Universidad Cat\'olica de Valpara\'\i so and Universidad Adolfo Ib\'a\~nez for hospitality, during the visit funded by CONICYT MEC 80150093. He is also thankful to the Physics Department of the University of Buenos Aires, where part of this work was done under the support of the Milstein program.
J.A.S.-G. acknowledges support from CUAASC grant of Chulalongkorn University and Spanish FPI fellowship from FEDER grant FPA-2011-22594.
A.V.L. is supported by the Spanish MECD fellowship FPU16/06675.

\appendix 
\section{T-duality rules and generic vielbein} 
\label{AppendixOtherVielbeins}

Our generalized Bergshoeff-de Roo action was originally found using so-called Double Field Theory (DFT); in particular, a vielbein formulation of it. For further details the reader is advised to go to  \cite{MarquesNunez} and references therein ---especially \cite{Aldazabal:2013sca} for a review of DFT in the $a=b=0$ case. In this formalism the fields live in a space of doubled number of dimensions, namely $2(D+1)$-dimensional in our case. For the purposes of this work we can understand DFT as a way to write the $(D+1)$-dimensional actions as manifestly invariant under T-duality. In fact, we derived the corrected rules of T-duality transformation using DFT. We are going to explain the procedure in this Appendix, and the frame \eqref{KMVielbein} will be particularly convenient. 

Let us briefly review some necessary rudiments of Double Field Theory. $2(D+1)$-dimensional fields are parameterized in terms of $D+1$ fields as explained in Section 3.4 of \cite{MarquesNunez}. There are two of them, namely the generalized DFT vielbein,
\begin{equation}
\mc E_{\cal M}{}^{\cal A} = \frac{1}{\sqrt{2}} \left(
\begin{array}{cc} \bar  E^{(+)}_A{}^M & - G^{AB} \bar  E^{(-)}_B{}^M \\ [0.5em]
\bar E^{(+)}_M{}^B G_{BA} -  \bar  E^{(+)}_A{}^R \bar  B_{RM} & \quad \bar E^{(-)}_M{}^A + G^{AB}  \bar  E^{(-)}_B{}^R \bar B_{RM} \end{array} \right) ~,
\label{GeneralizedVielbein}
\end{equation}
where $G_{A B}$ is the $(D+1)$-dimensional Minkowski metric, and the DFT dilaton, $e^{-2d} = e^{-2 \bar \Phi} \sqrt{-\bar  G}$. We denote with a bar those fields appearing in the DFT parametrization. Their properties under T-duality and the relation with the unbarred fields will be specified later in this Appendix. In DFT there is a symmetry that allows to rotate the two vielbeins $\bar E^{(\pm)}_M{}^A$ independently, with different Lorentz generators \cite{MarquesNunez}. In particular, a generalized infinitesimal transformation generated by
\begin{equation}
\Lm_{\cal A}{}^{\cal B} = \left( \begin{array}{cc}
\Lm^{(+) A}{}_B & 0 \\ [0.4em]
0 & \Lm^{(-)}\hskip-0.75mm_A{}^B
\end{array} \right) ~,
\label{DFTLorentz}
\end{equation}
induces the following transformation $\d_\Lm \mc E_{\mc M}{}^{\mc A} =  \mc E_{\cal M}{}^{\cal B} \Lm_{\cal B}{}^{\cal A} + \d'_\Lm \mc E_{\mc M}{}^{\mc A}$,
\begin{equation} 
\d_\Lm \mc E_{\mc M}{}^{\mc A} =  \mc E_{\cal M}{}^{\cal B} \Lm_{\cal B}{}^{\cal A} + \left( a\, \partial_{[\underline {\cal M}} \Lambda_{\mc C}{}^{\cal B}\, {\cal F}^{(-)}_{\overline {\cal N}] \cal B}{}^{\cal C} - b\, \partial_{[\overline{\cal M}} \Lambda_{{\cal C}}{}^{{\cal B}}\, {\cal F}^{(+)}_{\underline {\cal N}] {\cal B}}{}^{{\cal C}}\right) \cal E^{{\mc N }{\mc A}} ~,
\label{DFTAnomalousLorentzTransformation}
\end{equation}
where ${\cal F}^{(\pm)}$ are generalized fluxes whose specific form is not important for us, whereas underline and overline indices mean that either of the pair of complementary projections in double space was applied to the corresponding index (see \cite{MarquesNunez} for further details). One important case in which $\d'_\Lm \mc E_{\mc M}{}^{\mc A} = 0$ is when the transformation is uniform and then $\partial_{\mc M} \Lm_{\cal B}{}^{\cal A}$ vanishes; for such cases, $\d_\Lm \mc E_{\mc M}{}^{\mc A} =  \mc E_{\cal M}{}^{\cal B} \Lm_{\cal B}{}^{\cal A}$.

Notice from \eqref{GeneralizedVielbein} that \eqref{DFTAnomalousLorentzTransformation} involves non-trivial transformations in both $\bar E^{(\pm)}_M{}^A$ and $\bar B$, as well as in $\bar G_{MN}$ and $\bar\Phi$. The fact that Double Field Theory is the right framework to incorporate T-duality is transparent in the simple expression of the action written in this language:
\begin{equation}
\mathcal{I}_{\rm DFT} = \int dX e^{-2d} \left( {\cal R} - 2 \Lm  + a {\cal R}^{(-)} + b {\cal R}^{(+)} \right) ~,
\label{SDFTexplicit}
\end{equation}
where ${\cal R}$ and ${\cal R}^{(\pm)}$ are generalized diffeomorphism scalars whose explicit expressions can be found in \cite{MarquesNunez}. We can use the previous DFT symmetry transformation to obtain a unique vielbein, or, in other words, to make $\bar  E^{(-)}_M{}^A = \bar E^{(+)}_M{}^A$ through a gauge fixing condition. Once this is done, the DFT action can be rewritten as a $(D+1)$-dimensional theory with only one vielbein $\bar E_M{}^A \equiv \bar E^{(-)}_M{}^A = \bar   E^{(+)}_M{}^A$. Notice that the barred fields are \emph{not} the unbarred ones appearing in the rest of this paper, whose dynamics is described by the generalized Bergshoeff-de Roo action, $\mathcal{I}_{\rm BdR}$, given in \eqref{BergshoeffdeRoo}. Nevertheless, it can be shown that $\mathcal{I}_{\rm DFT} = \mathcal{I}_{\rm BdR}$ \cite{MarquesNunez} provided the following relations between barred and unbarred fields hold:
\begin{equation}  
(\bar E_M{}^A)^{(0)} = (E_M{}^A)^{(0)} ~, \qquad \bar B_{MN} = B_{MN} ~, \qquad e^{-2\bar\Phi} \sqrt{-\bar G} = e^{-2\Phi} \sqrt{-G} ~,
\label{BarredFieldRedefinition}
\end{equation}
whereas the only constraint in the choice of $(\bar E_M{}^A)^{(1)}$ comes from demanding that the corresponding $\bar G_{MN}$ fulfills
\begin{equation}  
\bar G_{MN} = G_{MN} - \frac{1}{4} \sum_{k=\pm} a_k\,\Omega_{MA}^{(k)~B}\,\Omega_{NB}^{(k)~A} ~.
\label{BarredMetricRedefinition}
\end{equation}
This is all we have to say about the relation between the DFT and the generalized Bergshoeff-de Roo actions. Now we will derive the T-duality rules for the latter starting from the former. The idea is to describe how the barred fields transform under T-duality. Once their transformation rules are known, we simply rewrite them in terms of the unbarred fields, finding in this way their $(a,b)$-corrected rules. This last step will be exemplified for $\wh G_{\psi\psi}$.

In Double Field Theory, the application of T-duality generates two different dual vielbeins even if before the duality $\bar E^{(-)}_M{}^A = \bar E^{(+)}_M{}^A$ in \eqref{GeneralizedVielbein} \cite{MarquesNunez}.\footnote{Indeed, even if $a=b=0$ in \eqref{BergshoeffdeRoo}, it is well-known that the T-dual of type IIA/B string theories contains two different dual vielbeins \cite{Hassan:1999bv}.} In particular, the two dual vielbeins $\wh{\bar E}{}^{(-)}_M{}^A$ and $\wh{\bar E}{}^{(+)}_M{}^A$ are given by:
\begin{equation}
\wh{\bar E}{}^{(\pm)}_\m{}^A = \bar E_\m{}^A - \frac{\bar Q^{(\mp)}_{\psi \m}}{\bar G_{\psi\psi}} \bar E_\psi{}^A ~, \quad\qquad
\wh{\bar E}{}^{(\pm)}_\psi{}^A = \mp \frac{\bar E_\psi{}^A}{\bar G_{\psi\psi}} ~,
\label{DFTVielbeinRules}
\end{equation}
where
\begin{equation}
\bar Q^{(\pm)}_{\psi \m} = \bar G_{\psi \m} \pm \bar B_{\psi \m} ~.
\end{equation}
Albeit not obvious at first glance, notice that both dual vielbeins lead to the same metric.\footnote{Transformations \eqref{DFTVielbeinRules} were also presented ---exchanging the names of the dual vielbeins; namely, our $\wh{\bar E}{}^{(+)}_M{}^A, \wh{\bar E}{}^{(-)}_M{}^A$ are equal to $\tilde   e_{(-) \m}{}^a, \tilde   e_{(+) \m}{}^a$--- in \cite{Hassan:1999bv}.} For the T-duals of $\bar B_{MN}$ and $\bar \Phi$, the rules are formally identical to the Buscher rules (but with barred fields). Now we have to use the freedom \eqref{DFTLorentz} to rotate one of the vielbeins so that they both become equal. This is explicitly achieved by means of a finite Lorentz transformation, $\bar\ml_B{}^A$, 
\begin{equation} 
\wh{\bar E}{}^{(-) A} = \wh{\bar E}{}^{(+)}{}^B \bar\ml_B{}^A ~, \qquad \bar\ml_B{}^A = \d_B{}^A - 2 \frac{\bar E_{\psi B} \bar E_\psi{}^A}{\bar G_{\psi\psi}} ~,
\label{LorentzForDuals}
\end{equation}
which satisfies $\bar\ml_C{}^B \bar\ml_B{}^A = \d_C{}^A$ and $\mathrm{det}\,\bar\ml_B{}^A = -1$ \cite{Hassan:1999bv}. We will refer to this procedure of equating one dual vielbein to the other via a Lorentz transformation as compensation. Because of the Double Field Theory generalized transformation rule \eqref{DFTAnomalousLorentzTransformation}, this compensation may induce changes in the other fields. On the other hand, a finite version of \eqref{AnomalousLorentzTransformation} is not available, and for this reason we do not derive the T-duality rules for a generic $(D+1)$-dimensional vielbein. This problem is solved, though, for a vielbein of the form \eqref{KMVielbein}, which has the nice property that $\bar E_M{}^A$ is also given by \eqref{KMVielbein}, thereby $\bar\ml_A{}^B =$ diag(${1,\ldots , 1, -1}$). As mentioned earlier, these uniform Lorentz transformations, $\partial_M \bar\ml_A{}^B = 0$, are symmetries of the full action in Double Field Theory and entail no anomalous modification of the fields. Consequently, the dual vielbeins in Double Field Theory are given by the rules \eqref{DFTVielbeinRules}, and we choose $\wh {\bar   E}_M{}^A = \wh{\bar E}{}^{(-)}_M{}^A$ as the dual vielbein in the $(D+1)$-dimensional theory written in barred fields.

We summarize the previous discussion by saying that no anomalous compensation is ever necessary in a frame of the form \eqref{KMVielbein}, and for this reason in the rest of this Appendix we work in such frame. As there is no anomalous compensation, the duals of $\bar B_{MN}$ and $\bar\Phi$ are simply:
\begin{equation}
\wh {\bar B}_{MN} = \wt {\bar B}_{MN} ~, \qquad e^{-2\wh {\bar\Phi}} \sqrt{-\wh {\bar G}} = e^{-2\bar\Phi} \sqrt{-\bar G} ~;
\label{DualsBPhi}
\end{equation}
that is, they are given by the standard Buscher rules. The same happens for the dual metric $\wh{\bar G}_{MN}$,
\begin{equation}
\wh {\bar G}_{MN} = \wt{ \bar G}_{MN} ~,
\end{equation}
which follows from the form of $\wt{\bar E}_M{}^A$. At this point, it only remains to relate the dual barred fields $\wh {\bar E}_M{}^A, \wh{\bar B}_{MN}, \wh {\bar\Phi}$, to the unbarred ones, $\wh E_M{}^A, \wh B_{MN}, \wh \Phi$, by means of \eqref{BarredFieldRedefinition}.

We just saw that $\wh {\bar G}_{M N}$, $\wh {\bar B}_{MN}$ and $\wh {\bar \Phi}$ are given by the uncorrected Buscher rules. This allows us to obtain the corrected T-duality rules for $\wh G_{M N}$, $\wh B_{MN}$ and $\wh \Phi$ displayed in \eqref{GeneralCorrectedRulesB}, \eqref{e2PhiG} and \eqref{GeneralCorrectedRulesMetric}. Let us point out how to compute $\wh G_{\psi\psi}$ to illustrate the procedure. Essentially, we begin with the standard Buscher rule for $\bar G_{\psi\psi}$,
\begin{equation}
\wh {\bar G}_{\psi\psi} = \frac{1}{\bar G_{\psi\psi}} ~,
\end{equation} 
and then substitute the barred fields in terms of $G_{MN}$, $B_{MN}$ and $\Phi$ according to \eqref{BarredFieldRedefinition},
\begin{equation}
\wh G_{\psi\psi} - \sum_{k=\pm} \frac{a_k}{4} \wh\Om^{(k)\, 2}_{\psi\psi} = \frac{1}{G_{\psi\psi}} + \sum_{k=\pm} \frac{ a_k \Om^{(k)\,2}_{\psi\psi}}{4\, G^2_{\psi\psi}} ~.
\end{equation} 
This expression is valid to linear order in $a$ and $b$, which allows us to substitute $a_k \wh \Om^{(k)\,2}_{\psi\psi}$ for $a_k \wt \Om^{(k)\,2}_{\psi\psi}$; regardless of the actual expression of the corrected rule for $\wh \Om^{(\pm)}_{M A}{}^B$, this substitution is valid because its leading order part will always be the same as $\wt \Om^{(\pm)}_{M A}{}^B$. This is all we need to rewrite the previous equation as
\begin{equation}
\wh G_{\psi\psi} = \,\wt G_{\psi\psi} + \sum_{k=\pm} \frac{a_k}{4} \prt{\wt\Om^{(k)\,2}_{\psi\psi} + \frac{\Om^{(k)\,2}_{\psi\psi}}{G^2_{\psi\psi}}} ~,
\end{equation}
where $\wt \Om^{(k)\,2}_{\psi\psi}$ can be readily obtained using \eqref{TorsionfulConnectionRulesSimplified}; this is the result presented in \eqref{GeneralCorrectedRulesMetric}. Using a similar procedure we derived the rest of the corrected T-duality transformation rules for $\wh G_{MN}$ and $\wh B_{MN}$. For the dilaton, we apply the following equalities:
\begin{equation}
e^{-2\wh \Phi} \sqrt{-\wh G} = e^{-2\wh{\bar  {\Phi}}} \sqrt{-\wh{\bar {G}}} = e^{-2 \bar\Phi} \sqrt{-\bar G} = e^{-2\Phi} \sqrt{-G} ~,
\end{equation}
where we have relied on the fact that the standard Buscher rules leave $e^{-2\bar\Phi} \sqrt{-\bar G}$ invariant.

\section{Detailed computation of the generalized Bergshoeff-de Roo entropy} 
\label{AppendixEntropyComputation}

In this Appendix we will explain in full detail the computations leading to the results in Section \ref{EntropyBdR}. After a brief review of our conventions for differential forms, we present the derivation of the entropy charge associated with each one of the three terms presented in \eqref{Action0}.

\subsection{Notation and conventions for differential forms}

Given a $(D+1-p)$-form $\tilde{F}$, we can consider $F$ to be the Hodge dual of $\tilde{F}$ with a change of sign and upper indices: $F^{N_1 \dots N_p} = - ( \star \tilde{F} )^{N_1 \dots N_p}$. Then:
\begin{equation}
\tilde{F} = \frac{1}{(D+1-p)!} \tilde{F}_{M_1 \dots M_{D+1-p}} \dd x^{M_1} \wedge \dots \wedge \dd x^{M_{D+1-p}} = F^{N_1 \dots N_p} (\dd^{D+1-p} x)_{N_1 \dots N_p} ~,
\end{equation}
where
\begin{equation}
(\dd^{D+1-p} x)_{N_1 \dots N_p} = \frac{1}{p! (D+1-p)!} \e_{N_1 \dots N_p M_1 \dots M_{D+1-p}} \dd x^{M_1} \wedge \dots \wedge \dd x^{M_{D+1-p}} ~,
\end{equation}
$\e$ being the volume $(D+1)$-form. As an example, for a Lagrangian $(D+1)$-form $\tilde{L} = \e\,\mathcal{L}$, we have that the \emph{dual} (in the previous sense, change of sign included) is $L = \mathcal{L}$ and then $\tilde{L} = \mathcal{L}\, \dd^{D+1} x$. This also shows that $\e = \dd^{D+1}\,x$. Another useful result is:
\begin{equation}
i_\zeta \tilde{L} = \mathcal{L}\, \zeta^M (\dd^{D} x)_M ~.
\label{DualInteriorProduct}
\end{equation}
Finally, under exterior differentiation we have:
\begin{equation}
\dd \tilde{F} = \nabla_P F^{N_1 \dots N_{p-1} P}\, (\dd^{D-p+2} x)_{N_1 \dots N_{p-1}} ~, 
\label{dualExteriorDerivative}
\end{equation}
so the dual to $\dd \tilde{F}$ is $\nabla_P F^{N_1 \dots N_{p-1} P}$ in this language. To compare this notation with the standard for differential forms, let us rewrite the defining equation of $Q_\G$, {\it i.e.}, $\dd Q_\G \cong j_\G$, which reads  $\nabla_N Q^{M N}_{\Gamma} \cong j^{M}_{\Gamma}$. If we can write the current vector, $j^{M}_{\Gamma}$, as a total derivative on-shell, we can immediately read the associated charge (up to innocuous ambiguities). Of course, if one manages to write $j$ as the exterior derivative of some codimension-2 form, one can achieve the same goal with differential forms. Nevertheless, in some particular calculations one notation is more convenient than the other. For this reason, we used tensors to compute $\mathcal I_0$'s charge, and differential forms in the case of the other two terms, $\mathcal I_{H'^2}$ and $\mathcal I_{R^2}$.

There is also an important result concerning the integration of differential forms. We will be mainly interested in integrating a $(D-1)$-form (the charge) on the bifurcation surface $\mathcal{B}$, so it would be useful to understand the form of $(\dd^{D-1} x)_{M N}$ when restricted to $\mathcal{B}$. As presented in more detail in \cite{Compere:2006my}, it can be shown that:
\begin{equation}
\left. (\dd^{D-1} x)_{MN} \right|_{\mathcal{B}} = \frac{1}{2} n_{MN}\,\bar{\e} ~,
\label{BifurcationSurfaceVolume}
\end{equation}
$\bar{\e}$ being the induced volume form on the bifurcation surface $\mathcal{B}$.

\subsection{Entropy charge of $\mathcal{I}_0$}

Consider now the Lagrangian form $L_0 = \e\,e^{-2\Phi} \mathcal{L}_0$ corresponding to the action $\mathcal{I}_0$ presented in \eqref{Action0}. Using the fact that, under a general variation of the fields, $\d \e = \frac{1}{2} G^{MN} \d G_{MN} \e$,
\begin{equation}
\begin{array}{l}
\d L_0 = \e \, e^{-2 \Phi} \left[ -2 \mathcal{L}_0 \, \d \Phi + 8 \nabla^M \Phi \nabla_M \d \Phi + \nabla_M X^M[\d G] \right. \\ [0.8em]
\quad\qquad\qquad\qquad \left. + \left( -R^{MN} - 4 \nabla^M \Phi \nabla^N \Phi + \frac{1}{2} G^{MN} \mathcal{L}_0 \right) \, \d G_{MN} \right] ~,
\end{array}
\end{equation}
where $X^M [\d G] = G^{PQ} \d\G^M_{PQ} - G^{MP}\d\G^Q_{PQ}$. Now, terms with $\d\Phi$ or $\d G_{MN}$ will be \emph{part} of the equations of motion (the other parts coming from $\mathcal{I}_{H'^2}$ and $\mathcal{I}_{R^2}$ in \eqref{Action0}). We can thus forget about them for the boundary term. To simplify the remaining terms, we have to take into account the symmetry transformations $\dLLz$ we employ to obtain the entropy charge. Considering both anomalous Lorentz invariance and the gauge symmetry of the $B$ field, the following are the Lie-anomalous Lorentz transformations:
\begin{equation}
\begin{array}{ccl}
\dLLz E^A & = & \LieDz E^A + E^B \lambda_B{}^A ~, \qquad \dLLz G_{MN} = \LieDz G_{MN} ~, \qquad \dLLz \Phi = \LieDz \Phi ~, \\ [0.8em]
\dLLz \Omega_A{}^B & = & \LieDz \Omega_A{}^B + \dd \lambda_A{}^B + \Omega_A{}^C \lambda_C{}^B - \lambda_A{}^C \Omega_C{}^B ~, \\ [0.7em]
\dLLz B & = & \LieDz B + \gm \Omega_A{}^B \wedge \dd \lambda_B{}^A + \displaystyle \frac{\gp}{2} \cH_{A}{}^B \wedge \dd\lambda_B{}^A + \dd \b ~, \\ [0.9em]
\dLLz \cH_A{}^B & \approx & \LieDz \cH_A{}^B - \lambda_A{}^C \cH_C{}^B + \cH_A{}^C \lambda_C{}^B ~,
\end{array}
\label{LieLorentzVariations}
\end{equation}
where $\Gamma = (\zeta, \lambda, \beta)$ are the transformation parameters. Recall that $\cH_{A}{}^B = H_{MA}{}^B\, dx^M$ \eqref{H1form}, and that $\approx$ means that we neglect $\mathcal{O}(a, b)$ terms. Now, in order to compute the entropy charge the first step is to find the charge $Q_\G$ in terms of generic transformation parameters $(\zeta, \l, \b)$. Then one simply substitutes those parameters by $(\xi, \lxie, \a_\xi)$, which make the previous variations to vanish on-shell. The reader should keep in mind that the charge has to be evaluated on the bifurcation surface $\bif$, and also that $\xi^M\inb = 0$. For this reason, terms in $Q_\xi$ which are linear in $\xi$ vanish at the bifurcation surface and will not contribute to the entropy integrand. Since we obtain the charge by doing two integrations by parts on $\d L$, the terms of $Q_\xi$ with $\nabla \xi$ come from those of $\d L$ with three or more derivatives. Other terms with less than three derivatives are not relevant for the entropy and will be ignored in the derivations that follow. A similar procedure was applied in \cite{Jacobson:1993vj}.

Based on this discussion, $\nabla_M \dLLx\Phi$ has at most two derivatives of the vector field, and is therefore irrelevant. However, the term with $X^M[\d G]$ will be relevant, and thus we are left with:
\begin{equation}
\d L_0 = \e\, e^{-2\Phi} \nabla_M X^M[\d G] + \ldots = \e\, \nabla_M \left( e^{-2\Phi} X^M[\d G] \right) + \ldots
\end{equation}
This is the relevant part of $\dd\theta_0(\Psi, \d\Psi)$, and using the dual notation introduced earlier we can easily read:\footnote{Up to the addition of a closed form to $\theta(\Psi, \d \Psi)$, which does not modify the entropy \cite{Jacobson:1993vj}.}
\begin{equation}
\begin{array}{ccl}
\theta^M_0 (\Psi, \d \Psi) & = & e^{-2 \Phi} \left(  G^{PQ} \d \G^M_{PQ} - G^{MP} \d \G^Q_{PQ} \right) + \ldots \\ [0.8em]
& = & 2 e^{-2 \Phi} G^{MN} G^{PQ} \nabla_{[P} \d G_{N]Q} + \ldots
\end{array}
\end{equation}
It is now a simple matter to construct the current $j^M_{0, \Gamma} = \theta_0^M(\Psi, \d_{\Gamma}\Psi) - \zeta^M e^{-2\Phi} \mathcal{L}_0$. Keeping only the relevant terms, it is given by
\begin{equation}
\begin{array}{ccl}
j^M_{0, \Gamma} &=& e^{-2\Phi} G^{MN} G^{PQ} \left( \nabla_P \nabla_{Q} \zeta_N - \nabla_N \nabla_{Q} \zeta_{P} \right) + \ldots \\ [0.8em]
&=& e^{-2\Phi} G^{MN} G^{PQ} \left( \nabla_P \nabla_{Q} \zeta_N - 2 \nabla_{[N} \nabla_{Q]} \zeta_{P} - \nabla_Q \nabla_N \zeta_P \right) + \ldots \\ [0.8em]
&=& e^{-2\Phi} \nabla_P \left( \nabla^P \zeta^M - \nabla^M \zeta^P \right) + \ldots \\ [0.8em]
&=& 2 \nabla_N \left( e^{-2\Phi} \nabla^{[N} \zeta^{M]} \right) + \ldots
\end{array}
\end{equation}
Notice the use of the Ricci identity in the second line to discard one of the terms. This is already in a suitable form to read the associated charge; using $\nabla_N Q^{MN}_{0, \Gamma} = j^{M}_{0, \Gamma}$, it is immediate to conclude:\footnote{Again, the primitive is defined up to closed form, but this ambiguity does not alter the entropy result \cite{Jacobson:1993vj}.}
\begin{equation} 
Q_{0, \Gamma} = - 2 e^{-2\Phi} \nabla^{M} \zeta^{N} \left( \dd^{D-1} x \right)_{MN} + \ldots
\label{ChargeTerm0}
\end{equation}
Defining $Q_0$ as $Q_{0,\G}$ for $\Gamma := (\zeta, \l, \b) = (\xi, \lxie,\a_\xi)$, we find the charge of $\mathcal I_0$ \eqref{Charge0BdR} presented in the main text.

\subsection{Entropy charge of $\mathcal{I}_{H'^2}$ and $\mathcal{I}_{R^2}$}

Let us start by analyzing the contribution of $\mathcal{I}_{H'^2}$. First of all, we need the following result for the general variation of a Chern-Simons form built out of a connection $\Omega$:
\begin{equation}
\delta\Theta = \frac{1}{3} R^A{}_B \wedge \delta \Omega^B{}_A - \frac{1}{6} \dd\left( \Omega^A{}_B \wedge \delta \Omega^B{}_A \right) ~.
\end{equation}
This is valid for any Lorentz connection, with or without torsion, and $R^A{}_B$ is the curvature 2-form associated with $\Omega$. In particular, the functional form of $\delta \Theta^{(\pm)}$ is exactly the same just including the appropriate superscripts $(\pm)$. The previous result allows us to write the variation of $H'$ after some algebraic manipulations as follows:
\begin{eqnarray}
\delta H' &=& \dd\delta B - 2 \gm R_A{}^B \wedge \delta\Omega_B{}^A - \gp R_A{}^B \wedge \delta\cH_B{}^A - \gp \dd\cH_A{}^B \wedge \delta\Omega_{B}{}^A \nn \\ [0.8em]
& & - \displaystyle \frac{1}{2} \gm \dd \cH_A{}^B \wedge \delta \cH_B{}^A - 2 \gp \Omega_A{}^C \wedge \cH_C{}^B \wedge \delta \Omega_B{}^A - \gm \Omega_A{}^C \wedge \cH_C{}^B \wedge \delta \cH_B{}^A \nn \\ [0.8em]
& & - \displaystyle \frac{1}{2} \gm \cH_A{}^C \wedge \cH_C{}^B \wedge \delta \Omega_B{}^A - \frac{1}{4} \gp \cH_A{}^C \wedge \cH_C{}^B \wedge \delta \cH_B{}^A \label{VariationBarH} \\ [0.7em]
& & + \dd \left[ \gm \Omega_A{}^B \wedge \delta \Omega_B{}^A + \displaystyle \frac{\gm}{4} \cH_A{}^B \wedge \delta \cH_B{}^A + \frac{\gp}{2} \Omega_A{}^B \wedge \delta \cH_B{}^A + \frac{\gp}{2} \cH_A{}^B \wedge \delta \Omega_B{}^A \right] ~. \nn
\end{eqnarray}
Two further results are needed in order to write down the general variation of our Lagrangian. The first follows from the Hodge dual definition,
\begin{equation}
\delta \star H'= \frac{1}{2} G^{MN} \delta G_{MN} \star H' + \star\,\delta H' ~,
\end{equation}
and the second is the identity $\star F \wedge G = \star G \wedge F$ for any pair of $p$-forms $F$ and $G$. Then, we obtain the full variation of $L_{H'^2} = -\half e^{-2\Phi}\star H' \wg H'$ as
\begin{equation}
\delta L_{H'^2} = e^{-2 \Phi} \left[ \delta\Phi - \frac{1}{4} G^{MN} \delta G_{MN} \right] \star H'\wedge H'- e^{-2\Phi} \star H' \wedge \delta H' ~,
\label{VariationLagrangianBR1}
\end{equation}
where $\delta H'$ is given by \eqref{VariationBarH}. Notice that the first term in the previous equation is going to contribute to the equations of motion without any further integration by parts and, therefore, $\theta_{H'^2}(\Psi, \d\Psi)$ will be obtained completely from the second term ---albeit not all of it is part of the boundary term, since it also contains contributions to the equations of motion. Now, there is an obstacle to apply the derivative counting argument presented in the previous subsection. It is correct for the part of the transformations depending explicitly on $\zeta$ since at the end of the calculation we are going to set $\zeta = \xi$ and evaluate at the bifurcation surface. It is also valid for the contribution proportional to $\lambda$, since we will evaluate for $\lambda = \lambda^E_\xi$, which is defined in \eqref{LieLorentzTransformationsExample} and contains a single derivative of $\xi$. But we cannot proceed in the same way with the gauge term $\dd \b$ appearing for the $B$ field. As a consequence, we will derive first the contributions to the entropy charge arising from $\zeta$ and $\lm$, leaving that of $\beta$ for later analysis.\footnote{To be as clear as possible with the following calculations, we write explicitly the parameters of the transformation we are considering instead of $\Gamma$. Parameters taken to be zero are not written.}

Suppose then for a moment that we are working with the symmetry transformations \eqref{LieLorentzVariations} without $\dd \beta$. As we said, all the contribution to the boundary term comes from the last part of \eqref{VariationLagrangianBR1}, and since $\delta H'$ is given by \eqref{VariationBarH} we can start our derivative counting process. First of all, in \eqref{LieLorentzVariations} we provided $\d_{\zeta, \lambda} \cH_{A}{}^B$ just to leading order, but this is enough given the form of \eqref{VariationBarH}; it is always multiplied by $\gp$ or $\gm$. Since $\lambda_{A}{}^{B}$ will have at most one derivative of the vector field when evaluated on $\lambda = \lambda^E_\xi$, its differential appearing in $\d_{\zeta, \lambda} B$ and $\d_{\zeta, \lambda} \Omega_A{}^B$ will have two derivatives. It is then easy to find the only terms containing three derivatives in $\delta L_{H'^2}$. After an integration by parts, these produce the following relevant part of the boundary term:
$$
\theta_{H'^2}(\Psi, \d \Psi) = (-1)^{D+1} e^{-2 \Phi} \Big[ \star H' \wedge \delta B + \star H \wedge \left( \gm \Omega_A{}^B + \frac{\gp}{2} \cH_A{}^B \right) \wedge \delta \Omega_B{}^A \Big] + \ldots ~,
$$
where we used the fact that $H' \approx H$. Now, in the current only terms containing two derivatives of $\zeta$ are relevant,
\begin{equation}
j_{H'^2, \zeta, \lambda} = (-1)^{D+1} e^{-2\Phi} \star H \wedge \left[ 2 \gm \Omega_{A}{}^{B} + \gp \cH_{A}{}^{B} \right] \wedge \dd \lambda_{B}{}^{A} + \ldots ~,
\end{equation}
and another integration by parts leads us to the charge presented in \eqref{ChargesHRBdR},
\begin{equation}
Q_{H'^2, \zeta, \lambda} = e^{-2\Phi} \star H \wedge \left[ 2 \gm \Omega_{A}{}^{B} + \gp \cH_{A}{}^{B} \right] \lambda_{B}{}^{A} + \ldots ~,
\end{equation}
after we take $\zeta = \xi$ and $\lambda = \lambda^E_\xi$. Similar calculations to the ones just presented allow us to obtain the contribution of $\mathcal{I}_{R^2}$; again, if we do not consider the gauge transformation term $\dd \beta$ in \eqref{LieLorentzVariations}. First of all, the variation of the Lagrangian is given by:
$$
\delta L_{R^2} = \sum_{k=\pm} \frac{a_k}{4} e^{-2 \Phi} \left[ \left( - \delta\Phi + \frac{1}{4} G^{MN} \delta G_{MN} \right) (\star R^{(k)}_A{}^B \wedge R^{(k)}_B{}^A) + 2 \star R^{(k)}_A{}^B \wedge \delta R^{(k)}_B{}^A \right] ~,
$$
where:
$$
\delta R^{(k)}_A{}^B = \dd \left( \delta \Omega^{(k)}_A{}^B \right) + \delta \left( \Omega^{(k)}_A{}^C \wedge \Omega^{(k)}_C{}^B \right) ~.
$$
Note that now all the relevant contribution to $\theta_{R^2}(\Psi, \d\Psi)$ will come from the first term containing the differential of $\delta \Omega^{(k)}_A{}^B$. It takes a simple calculation to conclude that
\begin{equation}
\theta_{R^2} (\Psi, \d \Psi) = (-1)^{D+1} \sum_{k=\pm} \frac{a_k}{2} e^{-2 \Phi} \star R^{(k)}_A{}^B \wedge \delta \Omega_B{}^A + \dots
\end{equation}
We can rewrite this expression in terms of the parameters $\gamma_{\pm}$ as
\begin{equation}
\begin{array}{ccl}
\theta_{R^2} (\Psi, \d \Psi) & = & - \displaystyle\frac{(-1)^{D+1}}{2} e^{-2\Phi} \Bigg\{ \star \left[ 4 \gp \left( R_A{}^B + \frac{1}{4} \cH_A{}^C \wedge \cH_C{}^B \right) \right. \\ [1.2em]
& & \left. \qquad\quad\quad\quad + 2 \gm \left( \dd \cH_A{}^B + 2 \Omega_A{}^C \wedge \cH_C{}^B \right) \right] \wedge \delta \Omega_B{}^A \Bigg\} + \dots
\end{array}
\end{equation}
The current is now given by
\begin{equation}
\begin{array}{ccl}
j_{R^2, \zeta, \lambda} & = & -(-1)^{D+1} e^{-2\Phi} \Bigg\{ 2 \gp \star \left( R_A{}^B + \displaystyle \frac{1}{4} \cH_A{}^C \wedge \cH_C{}^B \right) \wedge \dd \lambda_B{}^A \\ [1.2em]
& & \qquad\quad\quad\quad + \gm \star \left( \dd \cH_A{}^B + 2 \Omega_A{}^C \wedge \cH_C{}^B \right) \wedge \dd \lambda_B{}^A \Bigg\} + \dots ~~~,
\end{array}
\end{equation}
and the corresponding charge for the entropy would be
\begin{equation}
\begin{array}{l}
Q_{R^2, \zeta, \lambda} = - e^{-2\Phi} \Bigg[ 2 \gp \star \left( R_A{}^B + \displaystyle \frac{1}{4} \cH_A{}^C \wedge \cH_C{}^B \right) \qquad\quad\quad\quad \\ [1.2em]
\quad\qquad\qquad\qquad\qquad\qquad \gm \star \left( \dd \cH_A{}^B + 2 \Omega_A{}^C \wedge \cH_C{}^B \right) \Bigg] \lambda_B{}^A + \dots
\end{array}
\end{equation}
This is the result in \eqref{ChargesHRBdR} (taking $\zeta = \xi$ and $\lambda = \lambda^E_\xi$), but it is puzzling at first sight. We seem to have a $\gm$ contribution to the entropy, but Appendix B of \cite{MarquesNunez} shows that the action $\mathcal{I}_{R^2}$ in \eqref{Action0} has no $\gm$ part. Their proof relies upon Bianchi identities, and using them we can also conclude that the $\gm$ part of the entropy vanishes. Let us sketch the proof as follows. First of all, we can use the antisymmetry of $\lambda_B{}^A$ to rewrite:
$$
\star \left( \dd \cH_A{}^{B} + 2 \Omega_A{}^C \wedge \cH_C{}^{B} \right) \lambda_B{}^{A} = \star \left( \dd \cH_A{}^{B} + \Omega_A{}^C \wedge \cH_C{}^{B} + \cH_A{}^C \wedge \Omega_C{}^B \right) \lambda_B{}^{A} ~.
$$
This is a Lorentz covariant derivative for $\cH$,
\begin{equation}
\begin{array}{ccl}
Y^{AB} & := & \dd \cH^{AB} + \Omega^{AC} \wedge \cH_C{}^B + \cH^{AC} \wedge \Omega_C{}^B \\ [0.8em]
& = & (E^A)^M (E^B)^N \nabla_R H_{MNS} \, dx^R \wedge dx^S ~.
\end{array}
\end{equation}
This expression, when evaluated on $\mathcal{B}$, will be contracted with the binormal $n_{AB}$, since for $\lambda = \lambda^E_\xi$ we know that $\lambdaE_{BA}\inb = n_{AB}$. Besides, taking also the Hodge dual we obtain:
$$ 
\star (Y^{AB}n_{AB}) = \star \left(n^{MN} \nabla_R H_{MNS} \, dx^R \wedge dx^S \right) = 2 n^{MN} \nabla^{R} H^{S}{}_{MN} \left(\dd^{D-1} x\right)_{RS} ~.
$$
Using now $\left(\dd^{D-1} x\right)_{RS}\inb = n_{RS}\,\bar{\epsilon}/2$, we can show, as a consequence of $\dd H = 0$, that
\begin{equation}
n^{MN} n^{RS} \nabla_M H_{NRS} = \frac{1}{2} n^{MN} n^{RS} \left[ \nabla_{[M} H_{N]RS} + \nabla_{[R} H_{S]MN} \right] = 0 ~.
\end{equation}
So the $\gm$ terms in $Q_{R^2, \zeta,\l}$ vanish as they should. We finally obtain
\begin{equation}
S_{R^2} = - 4 \pi \gp \int_{\mathcal{B}} e^{-2 \Phi} \star \left( R^{AB} + \frac{1}{4} \cH^{AC} \wedge \cH_C{}^{B} \right) n_{AB} ~,
\end{equation}
which is the expression for the entropy presented in \eqref{entropyBR2}.

Let us now come back to the issue of the gauge symmetry of the $B$ field parametrized by $\b$. The first thing we have to realize is that these kind of gauge contributions to the entropy charge arise when considering both $\mathcal{I}_{H'^2}$ and $\mathcal{I}_{R^2}$. It will prove to be a good idea to tackle the full problem all at once, instead of isolating the two separate pieces. Consider then our full Lagrangian form $L = \mathcal{L} \, \e$, which depends on $B_{MN}$ only through $H_{MNR}$ and its first derivatives; the latter appearing from $R^{(\pm)}_{MNA}{}^B$ and $\Theta^{(\pm)}_{MNR}$. From a general variation just involving the $B$ field, it is easy to obtain
\begin{equation}
\begin{array}{ccl}
\d_B L \eq \e \left[ T^{MNR} \d_B H_{MNR} + S^{QMNR} \d_B \nabla_Q H_{MNR} \right] \\ [1.2em]
& = & - 3 \e \nabla_M \mathbb{E}^{MNR} \d B_{NR} + 3 \e \nabla_M \left[ \mathbb{E}^{MNR} \d B_{NR} + S^{M[QNR]} \nabla_Q \d B_{NR} \right] ~,
\end{array}
\end{equation}
where we have made use of the definitions \eqref{DefinitionsTandS}. The Euler-Lagrange equation for the $B$ field has the form:
\begin{equation}
\nabla_M \mathbb{E}^{MNR} \cong 0 ~,
\label{AbstractEoMB}
\end{equation}
whereas the boundary term is just
\begin{equation} \label{BoundaryTermB}
\theta^M (\Psi, \d B) = 3 \mathbb{E}^{MNR} \d B_{NR} + S^{M[QNR]} \d H_{QNR} ~.
\end{equation}
We can now easily obtain the contribution from the gauge parameter $\beta$ of the symmetry transformations \eqref{LieLorentzVariations}, that we denoted by $\d_\b$. Clearly, $\d_{\beta} H_{MNR} = 0$, and thus the contribution to the current and charge, proportional to $\beta$, will be:
\begin{equation}
j^M_{B, \beta} = 6 \mathbb{E}^{MNR} \nabla_N \b_R ~, \qquad Q^{MN}_{B, \beta} = 6 \mathbb{E}^{MNR} \b_R ~,
\end{equation}
where we employed the fact that $\nabla_N \mathbb{E}^{MNR} \cong 0$. This is \eqref{ChargeBBdR} if we define $Q_{\alpha_\xi} := Q_{B, \alpha_{\xi}}$. As a byproduct of this result, we can also conclude that the addition to $\alpha_{\xi}$ of an exact form will not change the entropy value, since taking $\alpha_{\xi} = \dd \g$ we can write $Q_{\alpha_{\xi}}$ as a total derivative to be integrated over the bifurcation surface, which we assume has no boundary, as in \cite{Jacobson:1993vj}.

\section{Stationarity of the corrected T-dual}
\label{AppendixStationarityOfTheDual}

In this Appendix we show the stationarity of the T-dual fields, namely that
$$
\d_\xi \wh E_M{}^A = \d_\xi \wh B_{MN} = \d_\xi \wh \Phi = 0 ~.
$$ 
It will follow automatically from $\li\Om^{(k)\,2}_{M N} \approx \li\wt\Om_{M N}^{(k)\,2} \approx 0$, so let us focus on this identity. We begin by noting that
\begin{eqnarray}
\d_\xi \Om_{M A}{}^B \eq \li \Om_{M A}{}^B + \mathcal D_M \lxieab = 0 ~,\\ [0.6em]
\d_\xi \Om^{(k)}_{M A}{}^B &\approx& \li \Om^{(k)}_{M A}{}^B + \mathcal D^{(k)}_M \lxieab \approx 0 ~,
\end{eqnarray}
which hold because $\d_\xi E_M{}^A= 0$ and $\d_\xi B = 0$. The first equality is taken from \cite{JacobsonMohd} and that of the second line is a consequence of \eqref{LieLorentzVariations} with $\zeta = \xi$ (the value of $\beta$ is irrelevant because $B_{M N}$ only appears through $H_{M N R}$). The operator $\mathcal D^{(k)}$ is defined as the Lorentz covariant exterior derivative $\mathcal D$ with $\Om^{(k)}$ instead of $\Om$. Since $\dd \lxieab = 0$, we can simplify the latter expression as
\begin{equation}
\li \Om^{(k)}_{M A}{}^B + \Om^{(k)}_{M A}{}^C  (\lxie)_C{}^B - (\lxie)_A{}^C \Om^{(k)}_{M C}{}^B \approx 0 ~.
\end{equation}
We see that the leading order effect of the Lie derivative on $\Om^{(k)}_{M A}{}^B$ is exactly a homogeneous Lorentz transformation with generator $-\lxieab$. From the previous equation one easily arrives to 
\begin{equation}
\li\Om^{(k)\,2}_{M N} \approx 0 ~.
\end{equation}
Let us address now the T-dual configuration. Since we want to repeat the argument above, we show first that $\d_\xi \widetilde E_M{}^A = 0$ and $\li \widetilde B = 0$. Indeed, under uncorrected Buscher rules \eqref{ReducedRules}, the components $E{}^a$ of a vielbein of the form \eqref{KMVielbein} are invariant, while $E^{\upsi}$ transforms into
\begin{equation}
\widetilde E^{\upsi} = e^{-\s}(W_\m \dd x^\m+\dd \psi) ~.
\end{equation}
Since $\li \s = 0$ and $\li W_\m= - \li B_{\psi\mu}= 0 $, it immediately follows that $\li \widetilde E^{\upsi} = 0$. The Lie derivatives act therefore on the Buscher-transformed vielbein the same way it does on the original one \eqref{LieVierbeinKruskal}:
\begin{equation}
\li \widetilde E_M{}^0 = \k \widetilde E_M{}^1 ~, \qquad \li\widetilde E_M{}^1 = \k \widetilde E_M{}^0 ~,
\end{equation}
while $\li\widetilde E_M{}^i = \li \widetilde E_M{}^{\underline{\psi}} = 0$. This means that $\delta_\xi \widetilde E_M{}^A  = \li \widetilde E_M{}^A  + \widetilde E_M{}^B (\l_\xi^{\widetilde E})_B{}^A= 0$, where the only independent non-vanishing component of $\l_\xi^{\widetilde E}$ is $(\l_\xi^{\widetilde E})_{ 0 1} = - \k$, and $\dd (\l_\xi^{\widetilde E})_A{}^B = 0$. Furthermore, $\li \wt B = 0$ because of \eqref{BuscherRulesB}. Taking $\wt \a_\xi=0$, we have the stationarity of $\wt B$, namely
$$
\d_\xi \wt B = \li \wt B +\frac 1 4 \big(b  \wt \Om^{(+)}_A{}^B -a \wt \Om^{(-)}_A{}^B \big)\wg \dd (\l_\xi^{\wt E})_B{}^A + \dd {\wt \a}_\xi = 0 ~.
$$
Therefore, we can repeat the reasoning applied before T-duality to conclude that $\li\widetilde \Om^{(k)\,2}_{MN} \approx 0$. Now that we have checked that $\li \Om^{(k)\,2}_{M N} \approx \li\widetilde \Om^{(k)\,2}_{M N} \approx 0$, it is easy to see in the corrected rules \eqref{GeneralCorrectedRulesMetric} that $\li \wh G_{MN} = \li \wh B_{MN} = \li \wh \Phi = 0$. This is enough to ensure the stationarity conditions taking $\wh \a_\xi = 0$ and using $\dd \l_\xi^{\wh E } \approx \dd \l_\xi^{\widetilde E} = 0$:
\begin{eqnarray}
\li \wh G_{M N} &=& 0 ~, \qquad \left( {\rm therefore} ~~\d_\xi \wh E_M{}^A = 0 \right) \\ [0.8em]
\d_\xi \wh B &=& \li \wh B +\frac 1 4 \prt{b  \wh \Om^{(+)}_A{}^B -a \wh \Om^{(-)}_A{}^B }\wg \dd (\l_\xi^{\wh E})_B{}^A + \dd {\wh \a}_\xi = 0 ~, \\ [0.8em]
\li e^{-2\wh \Phi} &=& \li \prt{e^{-2\Phi} \sqrt{-G}\, (-\wh G)^{-1/2}} = 0 ~. 
\end{eqnarray}
The implication between parenthesis in the first equation is discussed in reference \cite{JacobsonMohd}.

\section{Invariance of $e^{-2\Phi} \sqrt{G_h}$ under corrected T-duality}
\label{AppendixAreaInvariance}

In this Appendix we study the invariance of $e^{-2\Phi} \sqrt{G_h}$ under corrected T-duality, which plays an important role in Section \ref{SectionInvariance}. Let us start with the following property:
\begin{equation}
G_{\psi \bm}\inb = B_{\psi \bm}\inb =0~,
\end{equation}
where $\bm$ can be either $U$ or $V$. This metric component can be read from \eqref{GeneralKruskal} at $U=V=0$ (in fact in the whole horizon $V=0$). That of $B$ is derived from $\li B_{MN} = 0$. Then, it follows that:
\begin{equation}
\wt G_{\psi \bm} \inb = \wt G_{\a \bm} \inb = 0 ~, \qquad \wt G_{\bm \bn} \inb = G_{\bm \bn} \inb ~.
\end{equation}
We can use these results in the expression of the corrected T-dual fields \eqref{GeneralCorrectedRulesMetric}. Furthermore, we can resort to $\li \Om^{(k)\,2}_{MN} \approx \li\wt \Om^{(k)\,2}_{MN} \approx 0$ (this was shown in Appendix \ref{AppendixStationarityOfTheDual}) to make all $\Om^{(k)\,2}_{MN}$ components appearing in the expressions of $\wh G_{\psi \bm}\inb$ and $\wh G_{\a\bm}\inb$ vanish. Indeed, for any regular tensor $T_{M N}$:
\begin{equation}
\li T_{M N} \approx 0 \qquad\Rightarrow\qquad T_{\bm \alpha'}\inb \approx 0 ~.
\label{VanishingComponentsStationaryTensor}
\end{equation}
Notice that $\wt \Om^{(k)}_{MA}{}^B$ is regular because $\wt E_M{}^A$, $\wt E_A{}^M$ and $\wt B_{M N}$ are; see \eqref{BuscherRulesE}.\footnote{The reader should keep in mind that we always assume $G_{\psi \psi} \neq 0$, as mentioned at the end of Section \ref{SubsectionCoordinatesAndVielbein}. We also rely upon $e^{-2 \wh \Phi} \sqrt{-\wh G} = e^{-2\Phi} \sqrt{-G}$, which holds as well for $a=b=0$, and therefore $e^{-2 \wt \Phi} \sqrt{-\wt G} = e^{-2\Phi} \sqrt{-G}$. Using this expression and \eqref{BuscherRulesGmn}, one can prove that the determinant satisfies $\det \wt E_M{}^A = G^{-1}_{\psi \psi} \det E_M{}^A$, and then $\wt E_A{}^M$ is regular.} Thereby the desired property follows:  
\begin{equation}
\wt \Om^{(k)\,2}_{\bm \a'}\inb \approx \Om^{(k)\,2}_{\bm \a'}\inb \approx 0 ~.
\end{equation}
Substituting back in \eqref{GeneralCorrectedRulesMetric} we find that
\begin{equation}
\Gh_{\psi \bm} \inb = \Gh_{\a \bm} \inb = 0 ~, \qquad \Gh_{\bm \bn} \inb = G_{\bm \bn} \inb + \sum_{k=\pm} \frac{a_k}{4}\prt{\wt {\Om}^{(k)\,2}_{\bm \bn} -\Om^{(k)\,2}_{\bm \bn}} \Big\rvert_{\mathcal{B}} ~.
\label{SimplifiedDualMetric}
\end{equation}
The last two terms in $\wh G_{\bm \bn} \inb$ cancel each other. To see that this is the case, we convert curved $U,V$ indices to vielbein components $0,1$, taking into account that $E_M{}^0\inb$ and $E_M{}^1\inb$ are non-vanishing only when $M = \bm$ \eqref{VierbeinKruskal}. With a simple application of the dimensionally reduced T-duality rules \eqref{ReducedRules}, one arrives at
\begin{eqnarray}
a_k \wt {\Om}^{(k)\,2}_{\bm \bn} \inb \eq a_k \wt E_\bm {}^{\bar a} \wt E_\bn {}^{\bar b}\,\wt {\Om}^{(k)\,2}_{\bar a \bar b} \inb =  a_k E_\bm {}^{\bar a} E_\bn {}^{\bar b} \,\Om^{(k)\,2}_{\bar a \bar b} \inb ~, \\ [0.8em]
a_k \Om^  {(k)\,2}_{\bm \bn} \inb \eq a_k E_\bm {}^{\bar a} E_\bn{}^{\bar b} \,\Om^{(k)\,2}_{\bar a \bar b} \inb ~,
\end{eqnarray}
where $\bar a,\bar b$ can be either $0$ or $1$. Then:
\begin{equation}
\Gh_{U V} \inb = G_{U V} \inb ~, \qquad \Gh_{U U} \inb = 0 ~, \qquad \Gh_{V V} \inb = 0 ~.
\end{equation}
So in the end the corrected dual metric has a very simple block structure, and the components normal to the horizon turn out to be invariant under corrected T-duality:
\begin{eqnarray}
\wh G_{MN} \inb = \begin{pmatrix}
0 & G_{UV} \inb & 0 \\
G_{UV} \inb & 0 & 0 \\
0 & 0 & \wh G_{\a'\b'} \inb
\end{pmatrix} ~.
\end{eqnarray}
Notice that $\wh G_{UU} \inb = \wh G_{VV} \inb = \wh G_{\bm \a'} \inb = 0$ also follow from $\li \wh G_{MN} = 0$. Nevertheless, $\wh G_{UV}\inb$ and $\wh G_{\a'\b'}\inb$ are not constrained by it and, in fact, they do not vanish. This convenient block structure allows to establish:
\begin{equation}
e^{-2\wh\Phi} \sqrt{\wh G_h}\,\Big\rvert_{\mathcal{B}}= e^{-2\wh\Phi} \sqrt{-\wh G} \frac{1}{\sqrt{-\wh G_\perp}} \Bigg\rvert_{\mathcal{B}} = e^{-2\Phi} \sqrt{-G} \frac{1}{\sqrt{-G_\perp}}\bigg\rvert_{\mathcal{B}} = e^{-2\Phi} \sqrt{G_h} \Big\rvert_{\mathcal{B}} ~,
\end{equation}
and therefore the corrected invariance of the area law integrand of the entropy as presented in \eqref{ResultsAreaInvariance}. In the second equality we have used that $e^{-2\Phi} \sqrt{-G}$ is invariant under corrected T-duality \eqref{e2PhiG}.

\section{Independent check of the entropy formula}
\label{AppendixFieldRedefinition}

The purpose of this Appendix is to give an independent check of the entropy result \eqref{FullEntropyBR} when $\a_\xi = 0$, which is the case in Section \ref{SectionInvariance}. Let us start by introducing the generalized Metsaev-Tseytlin (MT) action \cite{MetsaevTseytlin}, which is equivalent to the generalized Bergshoeff-de Roo (BdR) action \eqref{BergshoeffdeRoo} using field redefinitions of $\mathcal{O}(a,b)$. The relation between the BdR and the MT fields is given by
\begin{equation}
\begin{array}{rcl}
G_{MN}|_{\rm MT} &=& G_{M N} - \displaystyle \frac{\gp}{2} H_M{}^{RS} H_{NSR} ~, \\ [0.9em]
B_{MN}|_{\rm MT} &=& B_{M N} + \gp \left( \nabla^R H_{RMN} - 2 \nabla_R \Phi H^R{}_{MN} - H_{[M}{}^{AB} \Omega_{N]}{}_{BA} \right) ~, \\ [0.9em]
\Phi|_{\rm MT} &=&  \Phi + \displaystyle \frac{\gp}{8} H_{MNR} H^{MNR} ~, 
\end{array}
\label{BRtoMTRedefinition}
\end{equation}
and the MT vielbein must also satisfy $E^{(0)}_M{}^A \vert_{\rm MT} = E^{(0)}_M{}^A$. Notice that the ambiguity in $E^{(1)}_M{}^A \vert_{\rm MT}$ is irrelevant as long as the metric is given by \eqref{BRtoMTRedefinition}. One can check this in the form of the MT action:
\begin{equation}
\mathcal{I}_{\rm MT} = \int\dd^{D+1}x \sqrt{-G}\, e^{-2\Phi} \big(L_{\rm MT}^{0} + L_{\rm MT}^{R^2} + L_{\rm MT}^{H'^2} \big) ~,
\label{MTAction}
\end{equation}
where
\begin{equation}
\begin{array}{ccl}
L_{\rm MT}^{0} &=& R - 2\Lm + 4 \nabla_M \Phi \nabla^M \Phi ~,\\ [0.9em]
L_{\rm MT}^{R^2} &=& \displaystyle\frac{\g_+}{2} \Big( R_{MNRS} R^{MNRS} - \frac{1}{2} H^{MNR} H_{MSL} R_{NR}{}^{SL} + \frac{1}{24} H^4 - \frac{1}{8} H^2_{MN} H^{2\, MN} \Big) ~, \\ [0.9em]
L_{\rm MT}^{H'^2} & = & - \displaystyle \frac{1}{12} \prt{H_{MNR} - 6 \gm \Theta_{MNR}}^2 ~,
\end{array}
\label{MetsaevTseytlin}
\end{equation}
and
\begin{equation}
H^4 = H^{MNR} H_{MS}{}^T H_{NT}{}^W H_{RW}{}^S ~, \qquad H^2_{MN} = H_{M}{}^{RS} H_{NSR} ~,
\end{equation}
all the fields in the previous equations being MT fields. We have added the cosmological constant term and also discarded a boundary term, that, as such, does not yield any entropy \cite{Jacobson:1993vj,Iyer:1994ys}; its precise expression is given in reference \cite{MarquesNunez}. The zeroth order part of the action is the same as in \eqref{BergshoeffdeRoo}. In $L^{0}_{\rm MT}$ we have also discarded the same boundary term as in the beginning of Section \ref{EntropyBdR}, as it does not contribute to the entropy. The first order in four-derivative corrections was obtained in \cite{MarquesNunez}, which encompass the results in \cite{MetsaevTseytlin}.

Let us now compute the entropy of the MT action \eqref{MTAction}. The contribution of $L_{\rm MT}^0$ and $L_{\rm MT}^{R^2}$ can be found with the standard Wald method \cite{Wald1993,Iyer:1994ys}. We have computed the entropy of $L_{\rm MT}^{H'^2}$ using the generalization of his method presented in Section \ref{SectionEntropy}.\footnote{A closely related action with anomalous diffeomorphism instead of anomalous Lorentz symmetry was studied in \cite{Tachikawa}, where it is assumed that the anomalous diffeomorphisms leave $B$ exactly invariant; therefore there is no term analog to $\dd\a_\xi$ \eqref{LieLorentzTransformationsExample} neither a contribution analogous to $Q_{\a_\xi}$ \eqref{ChargeBBdR} in \cite{Tachikawa}.} Adding the results for all terms in \eqref{MTAction} we arrive at:
\begin{equation}
\begin{array}{l}
S_{\rm MT} = 2\pi \displaystyle\int_\mathcal{B}\dd^{D-1}x e^{-2\Phi}\sqrt{G_h} \vert_{\rm MT} \Big[ 2 - \Big( \gp \Big( R^{MNRS} - \frac{1}{4} H^{TMN} H_T{}^{RS} \Big) \\ [1.2em]
\qquad\qquad\qquad\qquad\qquad\qquad\qquad\qquad\qquad - \gm H^{TMN} \Omega_T{}^{RS}\Big)\,n_{MN}\,n_{RS} \Big] ~,
\end{array}
\label{SMTappendix}
\end{equation}
where the leading order part and the terms proportional to $\gp$ and $\gm$ come respectively from $L_{\rm MT}^0$, $L_{\rm MT}^{R^2}$ and $L_{\rm MT}^{H'^2}$. At this point we have the form of the entropy of the generalized Metsaev-Tseytlin theory. The expression is already very similar to that of BdR \eqref{FullEntropyBR} for $\a_\xi = 0$, but the coefficient of the term $\gp H H$ is different.

The entropy integrals computed in both theories must have the same value \cite{Jacobson:1993vj}, namely $S_{\rm MT} = S_{\rm BdR}$. Therefore if we rewrite the MT fields in terms of the BdR ones we should arrive to \eqref{FullEntropyBR} with $\a_\xi = 0$. Notice, however, that the terms proportional to $\gp$ and $\gm$ in \eqref{SMTappendix} are already the same for both sets of fields to linear order in $a$ and $b$. Thereby, key contribution comes from the factor $e^{-2\Phi}\sqrt{G_h} \vert_{\rm MT}$. In the following this is checked explicitly.

It is convenient to work in the coordinates \eqref{GeneralKruskal}, in which the MT frame metric is of the form\footnote{There is a bifurcate Killing horizon on the MT metric, located in the same place as in the BdR metric. In general, a bifurcate Killing horizon is invariant under regular stationary field redefinition of the metric \cite{Jacobson:1993vj}.}
\begin{equation}
\dd s^2 \vert_{\rm MT} = G \prt{1 +\gp H^2_{UV}G^{-1}} \dd U \dd V + G_{\a' \b'}|_{\rm MT} \, \dd x^{\a'} \dd x^{\b'} ~ ,
\end{equation}
on the bifurcation surface. The leading order part of the binormal becomes
$$
n \inb=n_{U V} \dd U \wg \dd V \inb = \half G\,\dd U \wg \dd V \inb ~.
$$
Consequently, the entropy reads:
\begin{equation}
\begin{array}{l}
S_{\rm MT} = 2\pi \displaystyle\int_\mathcal{B}\dd^{D-1}x e^{-2\Phi}\sqrt{G_h} \vert_{\rm MT} \Big[ 2 + \gp \big( 4R^{UV}{}_{UV} - H^{ U V \a'} H_{U V \a'} \big) \\ [1.2em]
\qquad\qquad\qquad\qquad\qquad\qquad\qquad\qquad\qquad\qquad\qquad -4 \gm \Omega^{\a', U V} H_{\a' U V } \Big] ~.
\end{array}
\label{FullMTentropy}
\end{equation}
The next step is to rewrite $e^{-2\Phi}\sqrt{G_h} \vert_{\rm MT}$ in terms of BdR fields. To do so we use the invariance of $e^{-2\Phi} \sqrt{-G}$ under \eqref{BRtoMTRedefinition}, and the fact that on a horizon $\sqrt{-G} = \sqrt{-G_\perp}\sqrt{G_h}$, where $\sqrt{-G_\perp}$ is the volume orthogonal to the cross section. Using these two properties we have that, when evaluated on $\bif$,
\begin{equation}
\frac{e^{-2\Phi}\sqrt{G_h} \vert_{\rm MT} }{e^{-2\Phi}\sqrt{G_h} \vert_{\rm BdR}} = \frac{\sqrt{-G_\perp}\vert_{\rm BdR}}{\sqrt{-G_\perp}\vert_{\rm MT}} ~,
\end{equation} 
and we can use the identities
\begin{equation}
H^2_{UV} \inb \approx -G\, H^{U V \a'} H_{U V \a'} \inb ~, \qquad H^2_{\bm \a'} \inb \approx 0 ~,
\label{PropertiesHSquaredB}
\end{equation}
to conclude
\begin{equation}
\frac{e^{-2\Phi}\sqrt{G_h}\vert_{\rm MT}}{e^{-2\Phi}\sqrt{G_h} \vert_{\rm BdR}} =1-\gp  H^{UV \a'} H_{UV \a'} ~.
\end{equation}
The first identity in \eqref{PropertiesHSquaredB} follows from $\li H = \frac{1}{4} \dd\prt{a \Om^{(-)} - b  \Om^{(+)}} \wg \dd\lxieba \approx 0$ (obtained from $\d_\xi B = 0$) and a trivial generalization of \eqref{VanishingComponentsStationaryTensor} to a tensor with three indices. The second identity, instead, is a consequence of $\li H^2_{MN} \approx 0$ and \eqref{VanishingComponentsStationaryTensor}. The location of the bifurcate Killing horizon does not change under the field redefinition but the volume orthogonal to the bifurcation surface is not the same in this case. In the generalized Bergshoeff-de Roo frame $\sqrt{-G_\perp}\vert_{\rm BdR} = G/2$, and substituting back in the entropy formula,
\begin{equation}
\begin{array}{l}
S_{\rm MT} = 2\pi \displaystyle\int_\mathcal{B}\dd^{D-1}x e^{-2\Phi}\sqrt{G_h} \vert_{\rm BdR} \Big[ 2 + \gp \big( 4R^{UV}{}_{UV} - 3 H^{ U V \a'} H_{U V \a'} \big) \\ [1.2em]
\qquad\qquad\qquad\qquad\qquad\qquad\qquad\qquad\qquad\qquad\qquad -4 \gm \Omega^{\a', U V} H_{\a' U V } \Big] ~,
\end{array}
\label{fieldredefinitionmethodresultingeneral}
\end{equation}
which coincides with the generalized Bergshoeff-de Roo entropy \eqref{FullEntropyBR} for $\a_\xi=0$, this providing a quantitative check of our results.


\end{document}